\documentclass{mn2e}
\usepackage{hyperref}
\usepackage{psfig}
\usepackage{epsfig}
\usepackage{rotating}
\newif\ifAMStwofonts
\def\sqiglt{\hbox{\rlap{\lower.55ex \hbox {$\sim$}}\kern-.05em \raise.4ex \hbox{$<$}\,}}
\def\sqiggt{\hbox{\rlap{\lower.55ex \hbox {$\sim$}}\kern-.05em \raise.4ex \hbox{$>$}\,}}
\def\til{\ensuremath{\sim\,}}

\def\chisq{\ensuremath{\chi^2}}

\newcommand{\tim}[1]{\ensuremath{\times 10^{#1}}}

\def\etal{et al.\ }

\def\cms{\ensuremath{$cm$^{-2}}}
\def\cps{counts s$^{-1}$}

\def\swift{\emph{Swift}}

\def\nh{\ensuremath{N_{\rm H}}}
\def\arcsec{\ensuremath{^{\prime\prime}}}
\title[Swift-XRT GRB results]{Methods and results of an automatic analysis of a complete sample of
\swift-XRT observations of GRBs.}

\author[Evans et al.]{P.A. Evans$^1$\thanks{pae9@star.le.ac.uk}, A.P. Beardmore$^1$, K.L. Page$^1$, J.P. Osborne$^1$, P.T. O'Brien$^1$, 
\and R. Willingale$^1$, R.L.C. Starling$^1$,  D.N. Burrows$^2$, O. Godet$^1$, L. Vetere$^2$,
\and J. Racusin$^2$, M.R. Goad$^1$, K. Wiersema$^1$, L. Angelini$^3$, M. Capalbi$^4$, 
\and G. Chincarini$^{5,6}$, N. Gehrels$^3$, J.A. Kennea$^2$, R. Margutti$^{5,6}$, D.C. Morris$^{3,7}$,
\and C.J. Mountford$^1$, C. Pagani$^1$, M. Perri$^4$, P. Romano$^8$, N. Tanvir$^1$
\\
$^1$ X-ray and Observational Astronomy Group, Department of Physics and Astronomy, University of Leicester, 
LE1 7RH, UK
\\
$^2$ Department of Astronomy and Astrophysics, Pennsylvania State University,
525 Davey Lab, University Park, PA 16802, USA
\\
$^3$ NASA/Goddard Space Flight Center, Greenbelt, MD 20771, USA
\\
$^4$ ASI Science Data Center, ASDC c/o ESRIN, via G. Galilei, 00044 Frascati, Italy 
\\
$^5$ INAF, Osservatorio Astronomico di Brera Via E. Bianchi 46, I-23807,Merate (LC), Italy
\\
$^6$ Universit\'a degli Studi di Milano-Bicocca, Dipartimento di Fisica, Piazza
della Scienza 3, 20126, Milano, Italy
\\
$^7$ Center for Nuclear Studies, Department of Physics, The George Washington University, Washington,
DC 20052, USA
\\
$^8$ INAF, Istituto di Astrofisica Spaziale e Fisica Cosmica, Via U. La Malfa 153, I-90146 Palermo, Italy 
}

\date{Accepted 
      Received }

\pagerange{\pageref{firstpage}--\pageref{lastpage}}
\pubyear{2007}

\begin{document}

\maketitle

\label{firstpage}

\begin{abstract} 
We present a homogeneous X-ray analysis of all 318 Gamma Ray Bursts detected by
the X-ray Telescope on the \swift\ satellite up to 2008 July 23; this represents
the largest sample of X-ray GRB data published to date. In Sections 2--3 we
detail the methods which the \swift-XRT team has developed to produce the
enhanced positions, light curves, hardness ratios and spectra presented in this
paper. Software using these methods continues to create such products for all
new GRBs observed by the \swift-XRT. We also detail web-based tools allowing
users to create these products for any object observed by the XRT, not just
GRBs. In Sections 4--6 we present the results of our analysis of GRBs, including
probability distribution functions of the temporal and spectral properties of
the sample. We demonstrate evidence for a consistent underlying behaviour which
can produce a range of light curve morphologies, and attempt to interpret this
behaviour in the framework of external forward shock emission. We find several
difficulties, in particular that reconciliation of our data with the forward
shock model requires energy injection to continue for days to weeks.
\end{abstract}


\section{Introduction}
\label{sec:intro}

The \swift\ satellite (Gehrels \etal2004) has revolutionised our understanding
of Gamma Ray Bursts (GRBs), and the X-ray telescope (XRT, Burrows \etal2005) has
played a major role in this process. For example, XRT data have shown that the early
X-ray light curve is much more complex than originally thought, often
containing a shallow-decay `plateau' phase interpreted variously as energy
injection in a forward shock (Zhang \etal2006), reverse shock emission (Uhm \&\
Beloborodov 2007) and dust-scattering of the prompt GRB emission (Shao \&\ Dai
2007; although this has recently been ruled out by Shen \etal2008). The high
impact of the XRT has been made possible by \swift's unique rapid slewing
capability: the large field-of-view Burst Alert Telescope (BAT, Barthelmy
\etal2005a) detects GRBs, and \swift\ then automatically repoints itself so that
the narrow-field XRT and UV/Optical Telescope (UVOT; Roming \etal2005) can
observe the new burst within \til60--90 seconds of the trigger. The XRT has
detected \til95\% of \swift-BAT triggered GRBs, most of those within minutes of
the trigger, usually providing the most accurate rapidly available localisation
of the GRB and enabling early follow up with ground telescopes. \swift's X-ray
afterglow light curves and spectra, starting at typically $<$100 s have been
crucial to the discoveries made by \swift\ (see Zhang \etal2007; O'Brien
\etal2006; Willingale \etal2007, for reviews of the impact of \swift\ on GRB
science).

Because GRBs fade rapidly, follow-up observers need to make quick decisions
about whether or not to invest observing time on a given burst. This is
especially true of potentially unusual bursts, such as high-redshift candidates
or under-luminous GRBs. Thus it is desirable for GRB data to be rapidly
available and analysed quickly, reliably and ideally in a uniform manner.
\swift\ data are downlinked many times per day and are immediately processed by
the Swift Data Center (SDC) at Goddard Space Flight Center and made available to
the public via the SDC and data centres in the UK and Italy, minutes to hours
after the downlink. We (the \swift-XRT team) have developed software which, when
data arrive at the UK Swift Science Data Centre (UKSSDC), automatically
determines the best possible XRT GRB position and builds X-ray light curves and
spectra; the results are then published on the internet. This provides a
homogeneously generated catalogue of data, which we detail and discuss in this
paper (Section~\ref{sec:res}), following a description of the software. 

There are two types of GRB follow-up data telemetered from \swift: TDRSS and
Malindi data (described below). Initially, our software only worked with Malindi
data, however in 2008 February we modified our ground-based software to work
with TDRSS data as well (Evans \etal2008a); this provides better
positions, positions of fainter GRBs, and more reliable light curves and spectra
than those produced on board (although these should not be used for scientific
analysis).

\subsection{TDRSS data}

When \swift\ first detects and observes a GRB, some data are immediately
telemetered to the ground via NASA's Tracking and Data Relay Satellite System
(TDRSS). Among these `TDRSS data' are the XRT Single Pixel Event Report (`SPER')
data. \swift's XRT selects between Windowed Timing (WT, high time resolution but
only 1-D spatial information) and Photon Counting (PC, lower time resolution but
full spatial information) modes automatically, based on the count rate in the
central portion of the CCD (Hill \etal2005). When the XRT enters PC mode during
the first look at a new GRB, event lists containing every single-pixel (grade 0)
event above 0.55 keV detected within the central 200$\times$200 pixel region,
the SPER data, are delivered to the XRT team via the GCN system every 2
minutes until \swift\ slews away from the burst (up to \til2 ks after the
trigger). It must be noted that while we have made our products as reliable as
possible, SPER data are not fully calibrated and the light curves and spectra are
intended as quick-look products; they should not be used for scientific
analysis.

\subsection{Malindi data}

`Malindi data' are available several hours after a GRB trigger and comprise the
full observation dataset. Nine or ten times per day, \swift\ passes over the
Malindi ground station in Kenya and downlinks the data buffered on board. These
are then processed at the SDC and delivered to archives in the US, UK and Italy,
typically 90--120 minutes after the Malindi downlink. \swift\ also observes GRBs
which did not trigger the BAT, but are uploaded as Targets of Opportunity (ToOs).
For these bursts only Malindi data are telemetered.

Malindi data are grouped into \emph{observations}, usually one per day, each
with a unique ObsID. A single observation may contain many \emph{snapshots\/};
that is, pointings toward the source, since \swift\ is in a low Earth orbit and
thus its targets get occulted by the Earth once per orbit. The terminology
\emph{observations\/} and \emph{snapshots} is standard \swift\
parlance\footnote{http://heasarc.gsfc.nasa.gov/docs/swift/archive/archiveguide1/node3.html},
and will be used throughout this paper.

\subsection{GRB afterglow models}
\label{sec:fireball}

In this paper we introduce tools to produce high-precision positions, light
curves and spectra from XRT data, and describe the automatic application of these
tools to GRBs. This gives us a catalogue of results for all 318 GRBs detected by
the \swift-XRT up to GRB 080723B, and we discuss the implications for afterglow
science from this analysis. While our dataset can be used to test any models for
GRB emission, we do this in the conext of the fireball model (e.g.\ Rees \&\
M\'esz\'aros 1994, Sari, Piran \&\ Narayan 1998), which is the current consensus
model.

In this model the GRB progenitor launches highly relativistic jets of material
in a series of shells, of differing bulk Lorentz factors. Internal collisions
(i.e.\ within the jet) between shells cause shocks which radiate the GRB `prompt
emission'. The X-ray data presented in this paper may contain the tail of this
prompt emission, however it is thought to arise predominantly from the
afterglow. This is emission from an external shock which forms where the jet is
decelerated by the circumburst medium, and which propogates into that medium,
cooling by synchrotron radiation as it does so. See Piran (2005) for a
comprehensive review of the fireball model.

\subsection{Layout of the paper}
\label{sec:roadmap}

This paper is laid out thus:
\begin{itemize}
  \item{\S2 (pp \pageref{sec:prods}--\pageref{sec:ngrb}): XRT
  automatic analysis tools:
    \begin{itemize}
      \item{\S2.1 (pp \pageref{sec:spec}--\pageref{sec:astrom}): Spectra.}
      \item{\S2.2 (pp \pageref{sec:astrom}--\pageref{sec:lc}): Positions.}
      \item{\S2.3 (pp \pageref{sec:lc}--\pageref{sec:ngrb}): Light curves
      \&\ hardness ratios.}
    \end{itemize}
  }
  \item{\S3 (pp \pageref{sec:ngrb}--\pageref{sec:res}): Web tools
  to analyse any XRT source}
  \item{\S4 (pp \pageref{sec:res}--\pageref{sec:discuss}): Results of
  XRT GRB analysis}
  \item{\S5 (pp \pageref{sec:discuss}--\pageref{sec:interpret}): A
  canonical X-ray light curve?}
  \item{\S6 (pp \pageref{sec:interpret}--\pageref{sec:usage}):
  Understanding the X-ray afterglow}
\end{itemize}

Throughout this paper, errors quoted in all tables are at the 90\%\ confidence
level. Error bars on data in all figures are 1-$\sigma$ uncertainties.

\section{Automated data products}
\label{sec:prods}

We have developed software to produce three types of data products: `enhanced'
positions, light curves and spectra. The spectra were announced in Evans
\etal(2008b), and are presented in Section~\ref{sec:spec}. The method of
enhancing XRT positions has been previously documented (Goad \etal2007;
hereafter G07), however we have improved the algorithm, resulting in a factor of
\til2 improvement in precision (Section~\ref{sec:astrom}). The light curve code
has been published by Evans \etal(2007) and only minor modifications are described
here (Section~\ref{sec:lc}) along with details of new functionality which has
been made available to the user. We also describe automatic light curve fitting,
in Section~\ref{sec:fitlc}.

For GRBs which trigger the BAT, these products are created automatically while
GRBs observed as ToOs by \swift\ must be manually registered for automatic
analysis. This is usually done at the time of the ToO upload, so the data
products are available as rapidly as for BAT-detected bursts. Before any of
these products are produced, the XRT data are reprocessed at the UKSSDC
using the latest release of the {\sc xrtpipeline} tool\footnote{Part of the XRT
software, distributed with the HEASOFT package: 
http://heasarc.gsfc.nasa.gov/lheasoft/}. This
may differ from the version used at the SDC to create the cleaned event lists
available from the quick-look and archive sites. When a new version of the
\swift\ software or calibration is released, we carry out some tests to confirm
our product-generation code works reliably with the new release and then switch
to using the latest version. We do not however reprocess earlier GRBs with the
new release of the software.

\subsection{Spectra}
\label{sec:spec}

To create a spectrum from Malindi data, an image is formed from the first
available PC mode event list, and {\sc ximage} is used to identify any sources.
The brightest source within the BAT error circle (or equivalent for bursts
detected by other missions) is assumed to be the GRB, and the sky-coordinate
point-spread-function (PSF) fit routine developed for the position enhancement
(Section~\ref{sec:astrom}) is used to determine the source position in the XRT
astrometric frame. We use only observations  which begin within 12 hours of the
first one. Each observation is subdivided into snapshots, and these may be
further subdivided into times where pile up -- in which multiple photons are
registered as single events -- is or is not an issue. To identify intervals
affected by pile up we first search for times where the count rate within a 30
pixel radius circular region centred on the source is above 0.6 \cps\ in PC mode
or 150 \cps\ in WT mode. In PC mode we then obtain the PSF profile of the source
and compare it to the calibrated, non-piled-up PSF (Moretti \etal2005). This
indicates not only whether the source is piled up, but the radius out to which
pile up is a factor, $R_p$. If the source is piled up, we use an annular
extraction region to obtain source data, with an inner radius $R_p$. For WT mode
data we assume the data are piled up whenever the count rate exceeds 150 \cps\,
and use a box annulus extraction region, where the inner radius is that
necessary to keep the measured count rate below this level. For alternative
methods of identifying and eliminating pile up, see Vaughan \etal(2006) or
Romano \etal(2006). Where the data are not piled up, we use a circular
(rectangular) extraction region for PC (WT) mode. For each snapshot we determine
the mean source count rate and use this to choose the radius of the source
extraction region, according to Table~\ref{tab:sreg}; this maximises the
signal-to-noise ratio of the spectrum and is identical to the method used for
light curves.

Once the data have been divided into time intervals -- each with a source
extraction region -- we generate a source spectrum, source event list and full
frame event list for each one. Using the full frame event lists we create an
exposure map  and (using this and the source spectrum) an Ancillary Response
File (ARF) per interval. We then combine the source event lists from each time
interval using {\sc extractor} to get a single source spectrum. We similarly
combine the ARFs, using the {\sc addarf} tool; each ARF is weighted according to
the proportion of counts in the total source spectrum which came from this time
interval. This is not the same as extracting a single spectrum, exposure map and
ARF from an event list spanning multiple snapshots, since in that case the
weighting of the snapshots is determined by the exposure map (hence exposure in
each orbit), but GRBs are not of constant brightness. The ARFs must be correctly
weighted since the proximity of the source to the bad columns on the CCD, and
hence the effective detector area, will change from one snapshot to the next.
That weighting by counts is the correct approach is readily demonstrated: the
true number of counts from the the source ($C_t$) is simply the measured counts
($C_m$) multiplied by some correction factor ($Q$, implicit in the ARF) which
reflects counts lost to (for example) the bad columns. If there are multiple
snapshots, the true total number of counts is $C_t=\Sigma(C_m Q)=\Sigma C_m
\times \frac{\Sigma(C_m  Q)}{\Sigma C_m}$, i.e.\ a the overall correction factor
(or ARF) is the count-weighted mean of the individual ARFs.

The BACKSCAL keyword must be set for the source spectrum, for use by {\sc
xspec} (Arnaud 1996). Since this can be different for each time interval
included (the source region size is variable), this is taken as the weighted
mean of the BACKSCAL values from the individual source spectra extracted per
time interval, weighted according to the number of counts in those spectra.

A background spectrum is also produced. In WT mode background data are extracted
from the entire window, excluding a 120-pixel (283\arcsec) wide box centred on the
source. In PC mode the background region is an annulus with an inner radius of
60 pixels (142\arcsec) and an outer radius of 110 pixels (260\arcsec); if this extends
beyond the edge of the detector window it is shifted accordingly; the inner
circle of course remains centred on the source. To create the background
spectrum we do not subdivide the data more finely than the observations. For each
observation we identify any sources in the background region using {\sc ximage}
and exclude those areas from the extraction region. The individual observations'
background spectra are then combined as for the source spectra. 

The spectra thus produced were extensively verified by the \swift-XRT team. Our
test procedure consisted of producing spectra both manually and using this
software. These were both fitted with the same model and the best-fitting
parameters compared. We did this for more than thirty spectra and in every case,
the difference between the parameters from the automatic and manual spectra were
much less than the uncertainties on those parameters. We therefore conclude that
the automatic generation of spectra is reliable.

Spectra are automatically created for each new GRB observed by the XRT, and are
updated as new data are received. The results are posted online at
http://www.swift.ac.uk/xrt\_spectra in postscript and GIF format. We also
provide
a tar archive for download which contains the source and background spectra and
the ARF file necessary for users to fit the data themselves\footnote{Users also
require an RMF file, which can be found in the Calibration Database (CALDB,
http://swift.gsfc.nasa.gov/docs/heasarc/caldb/swift/).}. The results of automatic
spectral fitting (below) are available from the same website.

Since the creation of an XRT spectrum is useful for any target observed by \swift,
not just GRBs, we have created a tool to allow users to build spectra using our
software for any object observed by the XRT, see Section~\ref{sec:ngrb}.

\begin{table}
\begin{center}
\begin{tabular}{cc}
\hline
Count rate $R$ (\cps)  &  Source radius in pixels (\arcsec) \\
\hline
$R > 0.5$           & 30 ($70\farcs8$) \\
$0.1 < R \leq 0.5$  & 25 ($59\farcs0$)\\
$0.05 < R \leq 0.1$  & 20 ($47\farcs2$)\\
$0.01 < R \leq 0.05$  & 15 ($35\farcs4$)\\
$0.005 < R \leq 0.01$  & 12 ($28\farcs3$)\\
$0.001 < R \leq 0.005$  & 9 ($21\farcs2$)\\
$0.0005 < R \leq 0.001$  & 7 ($16\farcs5$)\\
$ R \leq 0.0005$  & 5 ($11\farcs8$)\\
\hline
\end{tabular}
\caption{Source extraction radii used for given PC-mode count rates. $R$ is the
measured, uncorrected count rate. This table is reproduced from Evans \etal(2007).
Values are given in XRT pixels and arcseconds: one XRT pixel corresponds to 2.36\arcsec.}
\label{tab:sreg}
\end{center}
\end{table}

\subsubsection{Automatic spectral fitting}
\label{sec:fitspec}

After creating the spectra, our software automatically models them with an
absorbed power-law.  Fitting is performed using the $X$-statistic in {\sc xspec
12}; we first apply a {\tt group min 1} command in {\sc grppha} (necessary for
{\sc xspec} to correctly calculate the \emph{C}-statistic, Arnaud, private
communication). A Response Matrix File (RMF) must also be used to fit the data;
the {\sc xrtmkarf} task used to create the ARF files when compiling the spectrum
selects the appropriate RMF from the Calibration Database (CALDB), and we supply
this file to {\sc xspec}. For a review of the XRT spectral response see Godet
\etal(2009).

The {\sc nh ftool}\footnote{http://heasarc.gsfc.nasa.gov/lheasoft/ftools} is
used to determine the Galactic column density in the direction of the burst
using the map from Kalberla \etal(2005), and the spectrum is fitted in with the
model {\tt phabs*phabs*pow}. The first absorption component is frozen at the
Galactic value and the second is free to vary. The abundances are fixed at those
from Anders \&\ Grevesse (1989). A tcl script ({\sc shakefit}) developed by
Simon Vaughan (Hurkett \etal2008; see their section~3.2.2), which uses the {\sc
error} command to detect and recover from local minima is then used to find the
true global best fit. If a spectroscopic redshift has been reported in the
literature, the XRT team can supply this and the second absorption component in
the model is replaced with a {\tt zphabs} component, with the redshift frozen at
that reported. The 90\%\ confidence intervals of each free parameter are found
using the {\sc error} command in {\sc xspec} which steps the parameter of
interest and repeats the fit until the \emph{C}-statistic has worsened by 2.706
compared to the best fitting value. The observed and unabsorbed 0.3--10 keV flux
for the model are also obtained from {\sc xspec}, and the former of these is
used to determine the count-rate to flux conversion factor for the GRB. This is
then automatically applied to the count-rate light curve to produce a flux units
version.

Very occasionally ($<1\%$\ of the time) {\sc xspec} finds a local minimum of the
\emph{C}-statistic rather than the best fit. This is usually immediately obvious from
the plot presented on the web pages, or because the best-fitting values are
unusual (the probability distributions in Section~\ref{sec:res} give a
quantitative definition of `unusual'). In this case a member of the XRT  team
will determine fitting parameters and supply these to the software,
which will then repeat the fit, using these parameters as the initial values.
When new data arrive and the spectrum is updated, the manually-supplied values
will again be used as the initial values for the fit. All of the results
presented in this paper have been verified by visual inspection, and the few
with poor fits corrected.

The results of these fits are presented in a table on the web page for each GRB,
accessible via http://www.swift.ac.uk/xrt\_spectra.

\subsubsection{Time resolved spectra}

Only time-averaged GRB spectra are produced automatically, however the software
allows for arbitrary time intervals to be specified, over which the spectra can
be compiled: on the web page presenting the spectra for each GRB is an option
to `Build time-sliced spectra'. Following this link users can specify times of
interest and their spectra will be built. If a member of the XRT team decides
that a particular time-resolved spectrum should be made available along with the
time-averaged spectra for a GRB they will add that spectrum to the main spectrum
results page of the burst. 

\subsubsection{Application to SPER data}

Spectra can also be obtained from the rapidly available SPER data using this
software. Note however that the SPER data have not been fully calibrated and
SPER spectra should be used as a quick-look product, rather than for scientific
analysis. Additional processing of the SPER data is necessary to prepare them
for spectral extraction. All SPER messages are combined into a single file and a
value of 100 is subtracted from the PHA (uncalibrated event energy) column,
(this was added on board to avoid negative values being obtained after bias
subtraction). A `Grade' column is then added (all events in SPER are grade 0).
The tool {\sc xrtcalcpi} is executed to create a PI (calibrated event energy)
column from which spectra can be built. At this point a spectrum can be
extracted. For SPER data we do not subdivide the data into intervals since they
only cover a single snapshot. If the data are piled up at any point an annular
source region is used for all of the SPER data. This means that we have only one
source spectrum and ARF. Fitting is performed as for Malindi data (see
Section~\ref{sec:fitspec}). Because SPER data only include events above 0.55
keV, it is harder to constrain the absorbing column.

\subsection{`Enhanced' positions}
\label{sec:astrom}

The technique of `enhancing' XRT positions from Malindi data by determining the
spacecraft attitude from images taken with the Ultra-violet/Optical Telescope
(UVOT) was first presented in G07. This technique produced
positions with error radii typically 40\%\ smaller than the `standard' positions
determined using the spacecraft attitude information determined from the
on-board star trackers. With all available data, the G07 positions were a
little less precise than those determined by Butler \etal(2007) using
serendipitous X-ray sources to find the correction; but while the latter
typically have large initial errors and only offer improvement over the
`standard' positions \til1 day after the trigger, the G07 positions were
available within hours of a trigger.

\begin{figure}
\begin{center}
\psfig{file=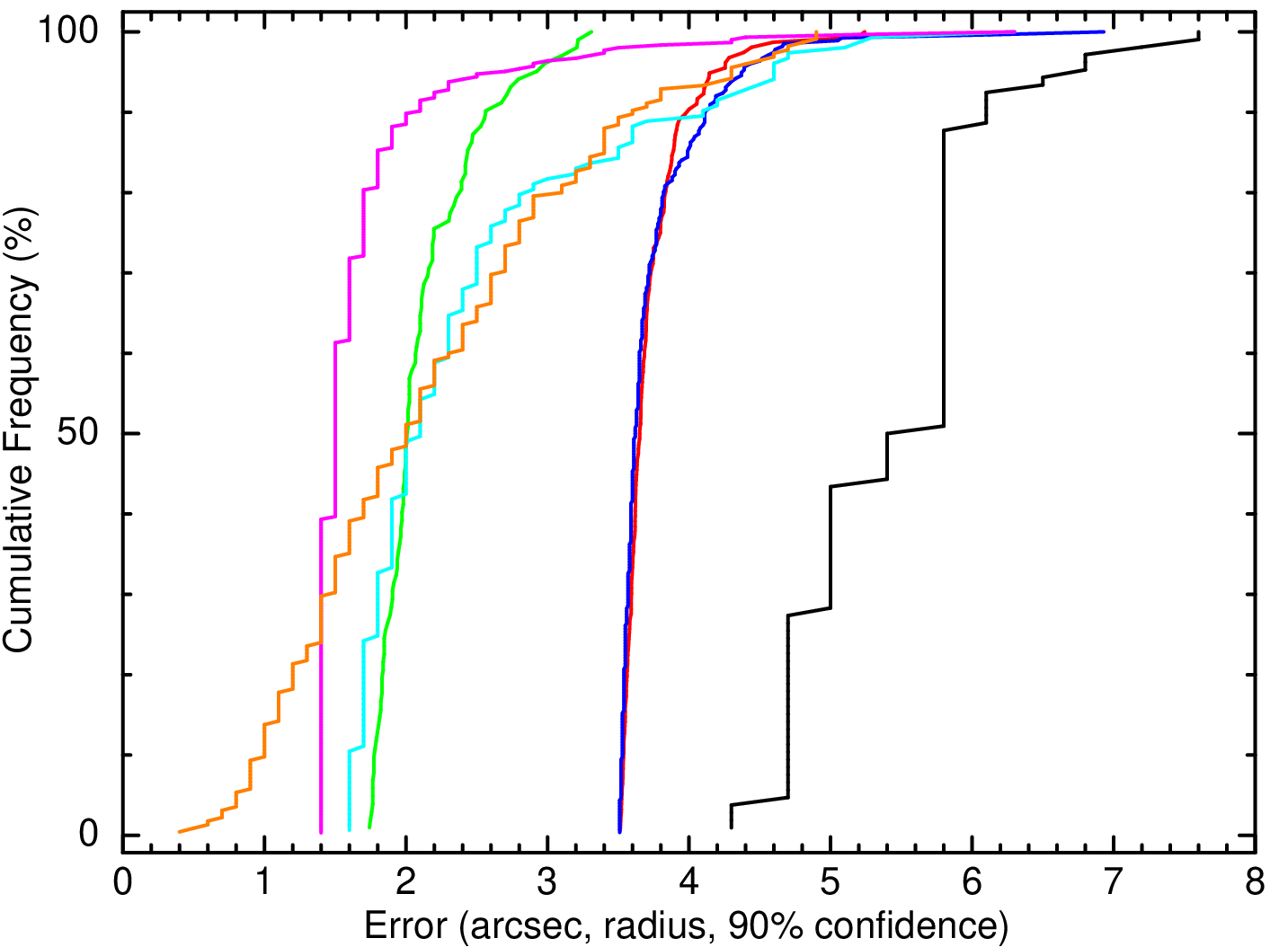,height=8.1cm,angle=-90}
\caption{The cumulative frequency of the 90\%\ confidence error radius
for XRT GRB positions determined using different techniques.
\newline \emph{Black}: on-board positions, \emph{Red}: positions from SPER data
derived using a PSF fit, \emph{Green}: enhanced SPER positions \emph{Blue}:
`refined' positions obtained from Malindi data using {\sc xrtcentroid},
\emph{Cyan}: enhanced Malindi positions using the G07 algorithm, \emph{Magenta}:
The enhanced Malindi positions using the algorithm presented here, \emph{Orange}
positions from Butler \etal(2007). For the latter, we have removed positions
with errors $>5\arcsec$ before calculating the distribution since these occur when
there is insufficient data to correct the position and bias the plot towards
larger uncertainties than are found when positions are successfully improved.}
\label{fig:errdist}
\end{center}
\end{figure}

We have made significant revisions to the enhancement process, reducing the
error radii by a further 30--50\%. These positions are still available
within hours of a trigger, and now in \til75\% of cases give better precision
than those of Butler \etal(2007), and are the most accurate XRT positions
available. Furthermore, we have developed a version of this algorithm applicable
to the SPER data, reducing the error radius of the prompt localisations by up to
60\%. These positions are typically available 10--20 minutes after a GRB
trigger. The relative precisions of the different XRT positions is shown in
Fig.~\ref{fig:errdist}.

Although we concentrate here on the improvements made since G07, it is
necessary to briefly summarise the algorithm used there, to give the context for
the improvements. For full details, see G07.

To produce an enhanced XRT position using the G07 method, the available data are
first split into `overlaps' -- times of simultaneous XRT (PC mode) and UVOT
($v$-band) data. For each such overlap a detector co-ordinate X-ray image is
produced and the GRB localised therein. This is done using a PSF-fit which
corrects for the effects of the bad columns on the CCD. The position is then
transformed into an equivalent position in UVOT detector co-ordinates, and
thence into UVOT sky co-ordinates, using the attitude information from the star
trackers on-board \swift. An image of the UVOT field of view is also
constructed, with the sky co-ordinates calculated using the same attitude
information. Serendipitous sources in the UVOT image are matched to the USNO-B1
catalogue, giving the quaternion needed to correct the image's attitude. This
quaternion is then applied to the previously calculated XRT GRB position in UVOT
sky co-ordinates, to give the `enhanced' XRT position for that overlap. This
process is performed for every available overlap, the weighted mean of all
overlaps is then calculated and systematic errors are added to give the enhanced
XRT position of the GRB. Since G07, this process has been improved in the
following ways:

1) The XRT-UVOT map, originally only derived for the UVOT $v$ filter, has been
extended to all UVOT filters. We have found, however, that limiting the software
to use just $v$, $b$ and white filters gave the best results\footnote{This is
probably because the USNO-B1 catalogue is an optical catalogue so matching UV
images to it is error prone.}.

2) Multiple observations can now be used, whereas originally only a single one
was used (recall that a single observation can still contain many snapshots). We
do not include all available observations as the X-ray afterglow will generally
be too faint to detect in a single overlap after \til12--24 hours and the
inclusion of more observations slows the enhancement process significantly.
Instead we use any observation which began within 12 hours of the first. Since
\swift's automatic response to a new GRB is a 60-ks observation, in many cases
this still means only the first observation is used and  this change to the code
then has no impact. It is not uncommon however, for the automatic observations
to be interrupted, either manually (to avoid overheating the XRT, Kennea
\etal2005) or automatically (i.e.\ \swift\ detects another GRB), in which case
multiple observations are suitable for use in the enhancement process.

3) We have substantially re-written the PSF fitting code used to determine the
position of the object on the XRT. The \emph{C}-statistic was previously calculated by
comparing the number of counts expected under the PSF with those in the data
(Cash 1978). It is now calculated by comparing the value of each pixel in the
image with that predicted by the model PSF. Additionally, the fit was previously
performed in XRT detector co-ordinates (so that the positions of the bad columns
on the CCD caused by the micrometeoroid impact, [Abbey \etal2005], were known),
but it is now performed in sky co-ordinates, using exposure maps to compensate
for bad columns or hot pixels.

These changes improve the enhanced positions for several reasons:

\begin{enumerate}
\item{Although \swift\ does not remain perfectly steady while on-source, the
star trackers provide precise \emph{relative} attitude information. Thus, while
the detector-coordinate image may be distorted by spacecraft motion, the sky
image will not be, and the PSF fit in this co-ordinate system will be more precise.}

\item{UVOT data are corrected for the spacecraft movement before being
downlinked. Every photon detected is shifted to the position it would have had
if the spacecraft had not moved since the image exposure began. As a result, the
effective time of the UVOT image is the start of the exposure. In contrast, a
detector-coordinate XRT image has no such correction and its effective time is
the mean photon arrival time within the image. Thus the aspect solution
determined from the UVOT image is actually for a different time, and possibly
slightly different attitude, than that at which the GRB position was measured on
the XRT detector. This was a contributor to the empirically-derived systematic
error in G07, see their section~6. XRT sky-coordinate images are
built using the sky-coordinate position of each event in the event lists which
were calculated on an event-by-event basis using the spacecraft attitude
information, and are thus analogous to the UVOT sky images. The sky co-ordinate
position of an object is therefore stable through a snapshot, allowing
the  {\sc pointxform} tool to be used to convert the measured sky-coordinate
position into XRT detector co-ordinates at the start of the UVOT exposure, so
the aspect solution and corrected position can be made contemporaneous. }

\item{Because the GRB position in XRT sky co-ordinates is stable within a
snapshot, we no longer need strictly simultaneous XRT and UVOT data. Instead,
we  use the longest XRT exposure possible without using the same data in
multiple overlaps, up to a maximum duration of an entire snapshot. For example,
if a snapshot contains two UVOT images, one in the $v$ filter and one in white,
the XRT data will be split in two -- the division occurring between the UVOT
exposures. An XRT position will be found for each part of the snapshot and
enhanced using the approximately contemporaneous UVOT data. On the other hand,
if the snapshot contains $v$ and uvm2 filter UVOT images, the XRT position will
be determined using the entire snapshot of data, and then enhanced using the $v$
image (we do not use the UV filter images to enhance positions, see above). This
means that overlaps can be used where the XRT source is fainter than was
possible in the previous version.}
\end{enumerate}

4) We have supplemented the improvement to the PSF fitting code by deducing PSF
profiles of piled up sources, in addition to the profile in the CALDB (Moretti
\etal2005). For each XRT image we perform the PSF fit 6 times; once with each PSF
profile. The fit with the lowest \emph{C}-statistic is then assumed to give the most
accurate position of the GRB.

To determine the piled-up PSF profiles we identified objects observed by \swift\
in PC mode which were piled up, away from the bad columns on the detector and
approximately constant in brightness. We then obtained the PSF profile using
{\sc ximage} and modelled it with a King$-$Gaussian function; the latter
component reflecting the counts lost to pile up. The centre of the Gaussian
profile was frozen at zero (i.e.\ the centre of the PSF). The PSF profiles thus
deduced are given in Table~\ref{tab:psf}, defined as King$(R_c,\beta)-N_{\rm
gaus}$Gaus($\sigma$), where $N_{\rm gaus}$ represents the relative normalisation
of the Gaussian component compared to the King component.

\begin{table}
\begin{center}
\begin{tabular}{ccccc}
\hline
Count rate & King $R_c$ & King $\beta$ & $N_{\rm gaus}$ & Gaus $\sigma$ \\
(\cps)     & (\arcsec)  &              &                & (\arcsec) \\
\hline
0.90    &  4.654   &  3.321 &  0.522  &  3.019 \\
1.36    &  6.578   &  3.627 &  0.504  &  4.139 \\
2.59    &  6.620   &  3.606 &  0.634  &  3.990 \\
5.15    &  12.790  &  3.505 &  1.039  &  6.411 \\
8.58    &  19.880  &  3.889 &  1.245  &  6.682 \\
\hline
\end{tabular}
\caption{PSF profiles for piled-up sources of different count rates. The profile
is defined as King$(R_c,\beta)-N_{\rm
gaus}$Gaus($\sigma$). }
\label{tab:psf}
\end{center}
\end{table}

5) \swift\ data are normally delivered to the data centres when the data from
all three instruments have been processed. However, during the first observation
of a new GRB special obsIDs are created which contain just the XRT or UVOT data.
These obsIDs -- which end with `991' for XRT or `992' for UVOT -- are usually
available before the instrument-combined dataset. We have modified our software
to use these special obsIDs, rather than waiting for the `all-in-one' obsID
data. This means that the positions are available up to half an hour earlier
than previously.

After implementing these changes, we ran the code for every GRB observed by
\swift, obtaining positions for 83\%. Following G07, we compared these
positions to the UVOT positions for those GRBs which were detected by that
instrument (taken from the GCN circulars). For each burst with an enhanced XRT
position and a UVOT position, we calculated the angular distance between these
positions, $R$. We also calculated the combined 90\%\ confidence error in the
enhanced XRT and UVOT positions $E_t=\sqrt(E_{\rm xrt\_enh}^2 + E_{\rm
uvot}^2)$. If $R/E_t\leq1$ the positions agree. If we included only the
statistical errors and the systematic uncertainty arising from the XRT-UVOT map
(1.3\arcsec), we found that fewer than 90\% of enhanced positions agreed with the UVOT
positions, as G07 did. However, we needed to increase our systematic only
marginally, to 1.36\arcsec, to achieve 90\% agreement, whereas in G07 we needed
a systematic of 1.5\arcsec. Our error radii are now  $\leq$1.5\arcsec\ 50\% of the time, and
$\leq$2.0\arcsec\ 90\% of the time. As Fig.~\ref{fig:errdist} shows, these represent the
best available XRT positions for the majority of GRBs.

The enhanced positions of all GRBs observed by \swift\ are available online at
http://www.swift.ac.uk/xrt\_positions. When a new GRB is  observed by \swift, an
enhanced position is automatically produced and added to this page as soon as
the necessary data are available. A GCN circular is also dispatched to advise
the community of the position, since in the majority of cases this represents
the best available position at this time and its rapid dissemination helps
follow-up astronomers. If no XRT counterpart was identified in the TDRSS data
this circular is not automatically dispatched. Under such circumstances, the
software enhances the position of the brightest source within the BAT error
circle; experience has shown that this is not always the GRB. Positions will
nonetheless be posted online even in these circumstances. As \swift\ continues
to observe the new GRB, the position is updated with each new delivery of data.
These updates appear at the above URL, but are not disseminated by other means.

Enhanced positions cannot be produced for all GRBs. If the burst is too faint to
be seen by XRT in a single snapshot the position cannot be found. Also,
sometimes the {\sc uvotskycorr} tool is unable to find sufficient matches
between UVOT field stars and USNO-B1 entries to return an aspect solution. Full
details of other, less common, reasons a position is not produced were given in
G07. To date 83\% of GRBs detected by the XRT have enhanced positions. If we
consider only BAT-triggered GRBs and exclude those where the UVOT was not in
operation, this number rises to 90\%.

Since many \swift\ non-GRB observations use the XRT to determine the position of
a newly-discovered (or known, but poorly localised) object, we have developed a
web tool to apply the position enhancement to user-selected non GRB objects. See
Section~\ref{sec:ngrb} for details.

\subsubsection{Application to SPER data}
\label{sec:sperastrom}

We can also apply the enhancement software to the SPER data. This was not
previously possible because the XRT and UVOT data had to be simultaneous -- for
the limited TDRSS data this is often not the case, or requires the XRT SPER data
to be filtered on time; given that SPER data contain substantially fewer events
than Malindi data this made determining the XRT position unreliable. Since the
modification to the PSF fit (point 3 above) removed the simultaneity
requirement, we can enhance positions using TDRSS data. The process is almost
identical to the Malindi data process described above, with the following
differences:

\begin{enumerate}
\item{Since we always have just one snapshot, the XRT position is determined once,
using all available data, and corrected multiple times; the statistical
uncertainty in the XRT position is thus only applied once, at the end of the
process.}
\item{We use the UVOT source lists telemetered via TDRSS, rather than the UVOT
images, to determine the aspect solution.}
\end{enumerate}

Point (ii) arises because the TDRSS source lists are calculated on
board from the full field-of-view image, however only a sub-image is downlinked;
thus the on-board source list provides more matches to the USNO-B1 catalogue and
a solution can be find more frequently.  The source localisation performed on
board is less sophisticated than is done on the ground, increasing the
systematic uncertainty needed to achieve 90\% agreement between our `enhanced
SPER' positions and the UVOT positions to 1.7\arcsec, compared to 1.36\arcsec\ for Malindi
data. In 2008 October the filter sequence followed by UVOT in its initial burst
response was changed to use the $U$ filter for the second exposure rather than
the White filter (the $U$ filter has a lower background rate and is thus
expected to find more afterglows than the White filter). We do not currently
have sufficient data to evaluate the effect of this change on the enhanced SPER
positions directly, however using historical Malindi data and $U$-filter images
taken later in the observations, we determined that our systematic error needed
to increase from 1.7\arcsec\ to 1.9\arcsec\ if the $U$ filter images were to be used. On the
other hand, not using the $U$ filter TDRSS data would mean most enhanced SPER
positions would be based on only one UVOT aspect solution and consequently have
errors typically 0.2\arcsec\ larger than hitherto. They would also be produced longer
after the burst as they would need to wait for the third UVOT exposure. Thus, on
2008 November 13 we modified the SPER position enhancement to use the $U$ filter
images and a 1.9\arcsec\ systematic error.

\subsection{Light curves and hardness ratios}
\label{sec:lc}

The software we used to generate automatic XRT light curves and hardness ratios
of GRBs was previously presented in Evans \etal(2007). This software has
received minor revision, which we describe briefly below, along with a significant
increase in functionality (Evans \etal2008b). 

\subsubsection{Modifications to the software}

1) Originally, the {\sc xrtcentroid} tool was used to determine the source position in
XRT co-ordinates. When the bad columns on the CCD intersected the source PSF,
this could result in an inaccurate position. This in turn could cause the
pile-up or bad column correction factors to be incorrect. We have altered the
software to use the PSF fit instead, since this gives accurate positions despite
the bad columns. This made small changes to a few light curves.

2) In the original version of the software, the final bin was always plotted as an
upper limit if it contained fewer than 15 counts (i.e.\ the errors could not be
considered Gaussian). In many cases the source is clearly detected at these
times and an upper limit is inappropriate. In 2007 June we modified the
software such that, if the final bin contains fewer than 15 counts, the Bayesian
approach of Kraft, Burrows \&\ Nousek (1991) is used to determine whether the
source is detected at the 3-$\sigma$ level. If it is, the Bayesian method is
then used to determine the 1-$\sigma$ confidence interval on the count rate, and
this is plotted as a point on the light curve. Otherwise an upper limit is
plotted as previously. Subsequent data deliveries are still included in this
final bin until it contains at least 15 counts, at which point the errors on the
bin are calculated using Gaussian statistics and a new bin is begun.

3) In 2008 July we fixed a minor bug in the software, which occasionally caused
Good Time Intervals (GTIs: times during which XRT was collecting data)
containing no events and lying between light curve bins to be ignored. This fix
made almost no difference to the light curves.

4) We made two other minor changes in 2008 July, of a cosmetic nature. First, we
amended the definition of the actual `time' value of a bin to be ten to the
power of the mean of the logarithms of the event times within the bin, rather
than the linear mean. This makes light curve plots reflect more accurately the
distribution of counts in a bin\footnote{For example, a bin with 4 events at
2\tim{4} s and a single event at 2\tim{6} s has a `time' of 3.2\tim{4} s using the
new method, but 4.2\tim{5} s using the standard mean.}. Second, at the end of an observing snapshot,
any events which are not yet sufficient to form a light curve bin are either
appended to the previous bin  or carried forward to the next bin, whichever
maximises the fractional exposure. Previously, these events could not be
appended to the previous bin if this meant spanning gaps in observations. This
change reduces the number of low fractional exposure bins in light curves. 

5) Now that we are producing automatic spectra of GRBs (Section~\ref{sec:spec}),
the counts-to-flux conversion factor determined from this is automatically used
to create a flux-units light curve. We still use a single conversion for the
entire light curve. This conversion factor is taken from the PC mode data
unless there are fewer than 200 events in the PC spectrum and more than 200 in
the WT spectrum, in which case the WT-mode conversion factor is used.

6) Four events lists are also now available for download: the source and
background event lists in WT and PC mode. As well as containing the events used
in the light curve, these contain the GTIs (which are used in the light curve
fitting, Section~\ref{sec:lcfit}) and the columns SRCRAD -- the radius of the
source extraction region, and PUPRAD -- the radius of the data excluded to
counter pile up. The former is in units of XRT pixels and the latter in
arcseconds, as required by the light curve software. One XRT pixel corresponds
to 2.36\arcsec.

7) Evans \etal(2007) stipulated that, if $C$ is the number of counts needed to
complete a bin in the main light curve, there should be $2C$ counts in each band
of the hardness ratio to complete a bin. This has been relaxed to $C$ counts in
each bin, giving significantly better time resolution in the hardness ratio.

As with the positions, we have also produced a web-based tool to create light
curves for non-GRB objects. This tool allows `conventional' binning, i.e. bins
of fixed duration, and is described in Section~\ref{sec:ngrb}.

\subsubsection{User-defined data binning}

In order to fully automate the light curve creation, the binning criteria
defined in Evans \etal(2007) are applied to all light curves, and these give a
useful, valid representation of the XRT data; but not necessarily the `best'
representation.

We have therefore produced a web-based tool to allow users to change the binning
criteria for a GRB. On the results page for each
GRB\footnote{http://www.swift.ac.uk/xrt\_curves/$<$target ID$>$} there is a link
entitled `rebin this GRB'. Following this link the user can specify the minimum
number of counts per bin in each XRT mode and whether this is to be used for all
bins or is `dynamic' (i.e.\ it varies with source brightness). On this page
there is a link to the `advanced' rebinning interface, which allows the user to
choose which event grades and energy bands are used in light curve creation. The
bands used for the hardness ratio can also be adjusted here. 

\subsubsection{SPER light curves}

This software can be easily applied to SPER data to produce light curves within
minutes of a trigger. These are of greater reliability than the light curves
telemetered via TDRSS, as the SPER light curves are background subtracted and
binned with the same method as used on Malindi data (Evans \etal2007). However,
SPER data were originally designed purely for source detection and localisation,
and thus do not contain GTI information; we are forced therefore to assume that
there is no dead-time in the light curve. Since SPER data only cover the first
snapshot this is a safe assumption unless the XRT switches back into WT
mode\footnote{This rarely occurs unless the GRB emits a strong flare, in which
case this is obvious from the SPER light curve.}.

\subsection{Automatic light curve fitting}
\label{sec:fitlc}

We have produced software to automatically fit broken power-law models to the
light curves. Note that this is not currently routinely applied to new data. The
fitting was performed using the least-squared approach implemented via the {\sc
Minuit2} minimisation routines produced at
CERN\footnote{http://project-mathlibs.web.cern.ch/project-mathlibs/sw/Minuit2/html/index.html},
using the \chisq\ statistic. Only light curves containing at least 3 bins were
fitted (i.e.\ only fits with at least 1 degree of freedom were attempted).

The procedure can be described as a four-step process:
\begin{enumerate}
\item{Identify `deviations' in the light curve}
\item{Ignore the times of flares, and fit a series of power-laws to the light
curve}
\item{Use the F-test to determine the best, justifiable fit}
\item{Check the results and repeat manually if necessary}
\end{enumerate}

These steps are discussed in detail below. Steps i--iii are automated, however
in step iv) a human can decide that the automatic results were incorrect, and
repeat those steps manually.

\subsubsection{Identifying `deviations'} \label{sec:flares} For this paper we
aim to characterise the light curves in terms of  power-law decays (see Nousek
\etal2006). However, many light curves show deviations from such behaviour which
must first be removed. The most common deviations are flares, but other
phenomena, such as the slow, curved rise seen in GRB 060218 (Campana \etal2006)
must also be ignored since they cannot be sensibly modelled as power-law decays.
To ensure that readers can compare model predictions with the \swift\ data
presented here, we list in Table~\ref{tab:flares} the times which were ignored
from the fits. This should not be considered a statistical sample of flares
since it includes other phenomena; also the times of the flare are based on when
the count rate in the light curve begins to rise (see below) rather than on
fitting a flare model to the light curve. For studies of flares in XRT light
curves, see for example Falcone \etal(2007); Chincarini \etal(2007); Kocevski,
Butler \&\ Bloom (2007).

Our automatic script to identify deviations first searches the light curve for
any interval where the count rate rises for at least two consecutive bins. If
this leads to an increase in count rate of at least $2\sigma$ a possible
deviation is deemed to have started.

To determine when this potential deviation ends we look for a shallowing of the
decay. We first identify the peak of the potential deviation as the highest
intensity bin before the count rate systematically drops again. Beyond this
peak, the software steps through each bin, calculating two decay indices for
each bin. If the current bin is bin $n$, these indices are:  $\alpha_1$: the
index of the decay from the peak of the deviation to bin $n$, and  $\alpha_2$:
the index of the decay from bin $n-1$ to bin $n$. A deviation is deemed to have
ended if at least one of the following conditions is met: a) There is a gap of
at least 1 ks between light curve bins, and prior to the gap the time was $<2$
ks since the trigger (i.e.\ they cannot begin in the first \swift\ snapshot
and extend into the second), or  b) Two out of three consecutive bins are found
with $\alpha_2<\alpha_1$ and both $\alpha_1$ and $\alpha_2$ decrease from one
bin to the next. 

The first condition exists because occasionally a flare starts towards the end
of the first \swift\ observing snapshot, and during the gap in observations the
flare ends and the light curve enters its final decay; thus the end of the flare
using the second test is never found. Chincarini \etal(2007) found that flares
tend to have duration:midpoint ratio of \til0.1, so we allow later time flares to
span snapshots.

Any time identified above is only confirmed as a genuine deviation, and hence
ignored for power-law fitting, if it contains fewer than 10 changes in XRT mode
(so is not an artefact of mode switching), contains at least 2\%\ of its
component bins before the peak and has a `significance' (defined below) of 1.8.
If the potential deviation occurs in the first snapshot (i.e.\ within 2 ks of
the trigger), the `significance' must be at least 3 for it to be confirmed.
These numbers were arrived at by trial and improvement to minimise the numbers
of false positives and negatives. The significance of a deviation is defined as
the peak count rate minus the pre-deviation count rate, divided by the errors in
these values added in quadrature.

\subsubsection{Fitting power laws}
\label{sec:lcfit}

Once the deviations have been identified and removed, we fitted the remaining
count-rate light curve with power-law segments separated by zero to five breaks.
Note that these breaks are not smoothed. Upper limits were excluded from the
fit. To fit the light curves correctly is non-trivial because of two,
interconnected factors: the bins have a finite duration, and many bins have a
fractional exposure less than 1. The normal way to calculate \chisq\ --
comparing model and data at the `time' of the bin -- does not account for the
finite (and sometimes very large) duration of the bin and the evolution of the
light curve through this. Instead, one should compare the number of counts
detected in each bin to the number predicted by the model. Determining the
number of counts predicted by the model in a given bin is non-trivial. The
simplest approach (method 1) is to integrate the model across the light curve
bin. However if the fractional exposure in the bin is not unity this technique
fails, since the model has been integrated over a longer time interval than that
during which counts were being collected. Renormalising the model by the
fractional exposure is not a valid solution: this assumes that the source
count-rate is the same during the dead-times and live-times within the light
curve bin; for GRBs, which fade, this is clearly untrue. 

A better fitting technique (method 2) is to use the Good Time Interval (GTI)
information available with the light curves\footnote{This is contained in the
source event lists, available from the \swift\ light curve repository.} and to
integrate the model across the times during which the XRT was collecting data.
Unfortunately, it has not been uncommon for the XRT to `mode switch' (to toggle
rapidly between PC and WT modes). This creates a large number of short GTIs.
Integrating the model over each of these GTIs dramatically slows down the fit,
especially for well-observed bursts, rendering this method impractical. Although
recent software and operational changes mean that mode-switching is now a rare
occurrence, it has been sufficiently common in the past that to fit all XRT
light curves using method 2 would take several months.

\begin{figure}
\begin{center}
\psfig{file=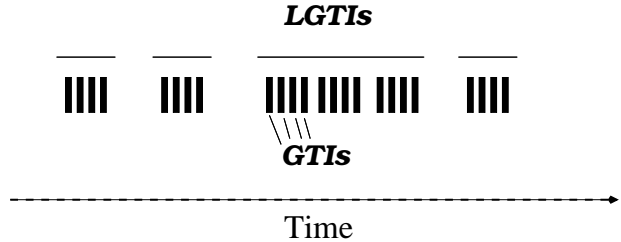,width=8.1cm}
\caption{An illustration of the long good time interval (LGTI) concept. The
vertical rectangles represent individual GTIs. The lines above show the LGTIs --
groups of GTIs with less than 30 s of dead-time between consecutive intervals.}
\label{fig:lgti}
\end{center}
\end{figure}

We thus tried a compromise (method 3), whereby within a light curve bin we group
GTIs together into `long GTIs' (LGTIs). An LGTI is defined as a cluster of GTIs
with less than 30 s of dead-time between consecutive GTIs. This is illustrated
in Fig.~\ref{fig:lgti}. The model value for a given bin is found by integrating
the model across each LGTI, multiplying by the fractional exposure in the LGTI
to correct for dead-time and then summing this integration for all LGTIs within
the light curve bin. This method, like method 1, contains an simplifying
assumption; in this case, that during the dead-time within an LGTI the model can
be assumed to be constant and at its mean value for the LGTI. This is a much
more defensible assumption from method 3 than that for method 1; The dead-time
within an LGTI is generally very short compared to the duration  of the LGTI,
whereas the deadtime in a bin can be a substantial fraction of the bin duration.
To check whether this assumption affects the fitted models, we chose 6 GRBs
containing bins of non-unity fractional exposure, and modelled their light
curves using both methods 2 and 3. We found that almost all the fitted
parameters and errors agreed to at least three significant figures; where they
only agreed to two, the parameter errors were large.

For each light curve our automatic script first fitted an unbroken power-law and
then added breaks and refitted with up to a maximum of five breaks (fewer if there
was less than one degree of freedom before this). Although break time is a free
parameter, it was necessary to estimate the time at which a break would most
improve the \chisq\ before adding it, to reduce the liklihood of the fit
converging on a local minimum. To achieve this, the software compared the data
with the previous model and identified time intervals where the data lie
systematically above or below the model. It calculated the \chisq\ contribution
from each such interval  and added the break at the end of the interval with
the greatest \chisq\ contribution, or the start of the interval if it extended
to the end of the observation.

\subsubsection{F-test}

For each broken power-law fit performed in the previous step, \chisq\ is
compared to that from the fit with one fewer break, and an F-test used to
determine whether the break is significant. We define a break as significant if
the F-test returns a probability of the \chisq\ improvement as  $<0.3$\%. Note
that we don't interpret this quantitatively as confirming the break at the
$3\sigma$ level, rather we use it as a convenient means of determining how many
breaks to use. Even if the break is not deemed significant in this way, the
software still adds a further break; sometimes the `true' best fit requires 2
breaks, but the \chisq\ improvement from a no-break fit to a one-break fit is
not significant.  The fit with the most breaks which is deemed significant by
the test above is taken as the `best' fit.

\subsubsection{Human intervention}

For each GRB, we checked the results of the automated steps above. In 23\%\ of
cases flares were misidentified; occasionally genuine flares were missed, but
most of the failures were false positives. In these cases we manually
defined the times to be excluded and re-ran the automatic power-law fitting.

\begin{figure*}
\begin{center}
\psfig{file=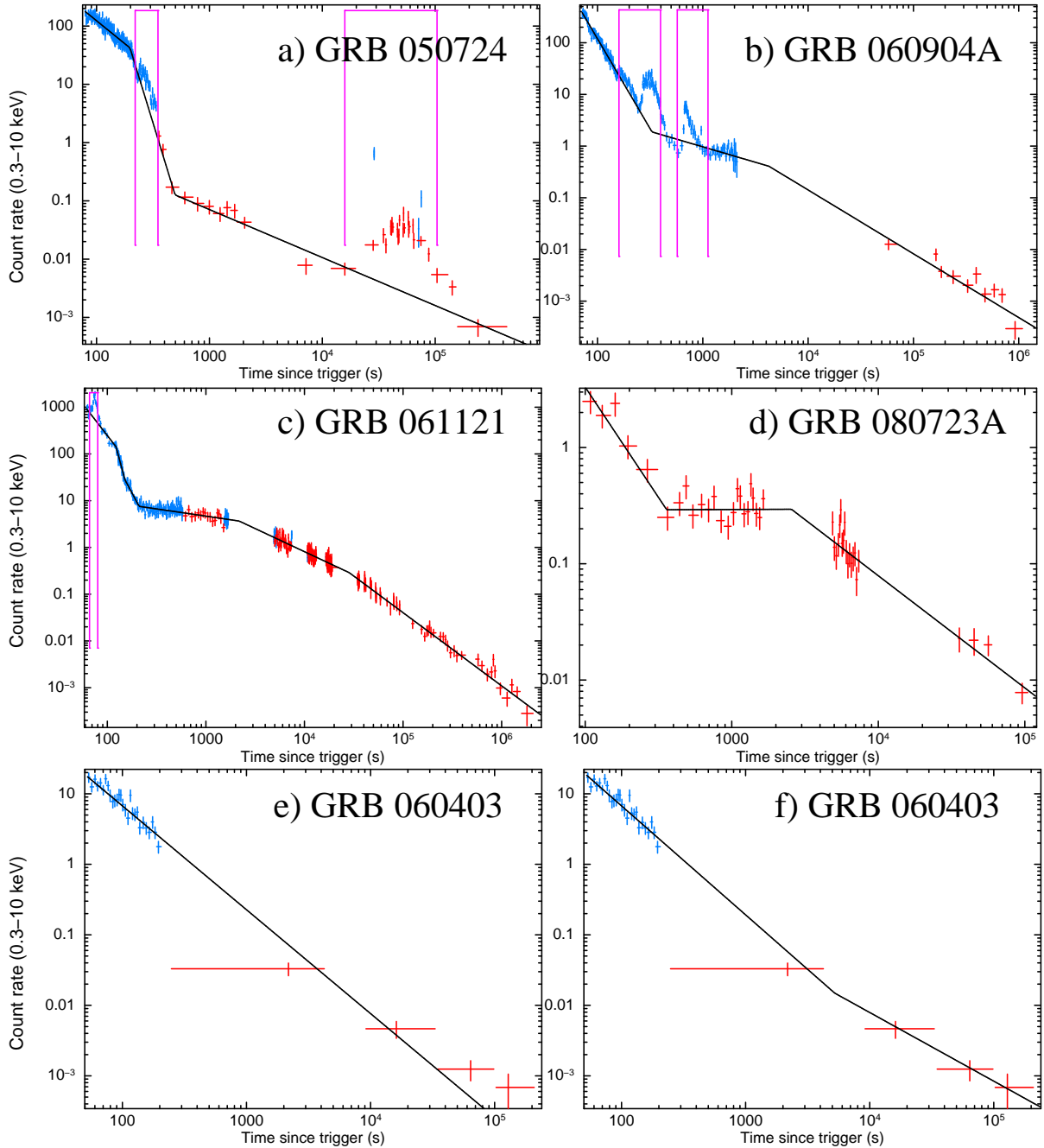,width=16cm}
\caption{Examples of automatically fitted light curves. WT data are shown in
blue, PC data in red. The purple boxes mark times identified as deviations from
power-law behaviour and excluded from the fit. Panels a)-d) show cases where no
human intervention was needed. Panels e) and f) illustrate a case where we
overruled the automatic determination of the `best' model. The break added in f)
is only significant at the 96\% level according to an F-test, so the automatic
software suggested e) as the best fit. However, given that we know that GRB
light curves tend to flatten in the first few kiloseconds, we believe that the
automatic fit with a break shown in f) is the most appropriate fit.}
\label{fig:autofit}
\end{center}
\end{figure*}

In \til5\%\ of cases visual inspection suggested that use of the F-test had not
identified the true best fit. Sometimes this was because one of the fits had
found a local minimum of \chisq\ rather than the best fit. In these cases we
manually adjusted the parameters and refitted until the true best fit was found
and re-performed the F-test. In other cases (\til1\%\ of light curves), the
F-test  deemed a break necessary only at the 90--99\% level, however, knowing
that light curves often show a `steep-shallow-steep' behaviour (Nousek
\etal2006; Zhang \etal2006) allowed us to confirm that the break was genuine --
an example of this is given in Fig.~\ref{fig:autofit}, along with several
examples of good automatic fits.

\section{Non-GRBs}
\label{sec:ngrb}

So far we have only discussed GRBs, since it is for these that we create
products automatically, and  present results below. However, \swift\ regularly
observes other types of source, with the XRT data often being crucial to the
science goals of the observation. We have thus adapted the three tools discussed
in Section~\ref{sec:prods} for use with non-GRB sources, and created a web
interface to execute them on-demand. This is  available at
http://www.swift.ac.uk/user\_products. In this section we briefly detail the
differences between the `non-GRB' versions of the software and those described
in Section~\ref{sec:prods}. While these tools are public, we refer users to the
usage policy at the end of this paper.

\subsection{Spectra}

The spectrum creation and fitting software for non-GRBs is almost identical to
that used for GRBs, except that the user can choose which observations are used
to form the spectrum. The user can also specify up to 4 time intervals rather
than creating a single time-averaged spectrum (for GRBs this is possible as
well, but only after the average spectrum has been automatically produced). As
with GRBs, the spectra are automatically fitted with absorbed power-law models,
and the spectral files are provided for download so users can fit other models
and interact with the data as required.

\subsection{Positions}

As noted in Section~\ref{sec:astrom}, if UVOT images taken in the UV filters are
used to enhance XRT positions, we tend to find larger errors, thus for GRBs we
use only the $v$, $b$ and white filters. However, many non-GRB observations are
performed using the `filter of the day' (to prolong the life of \swift's filter
wheel), which is usually one of the UV filters. Thus the non-GRB position
enhancement tool works in two passes: first it tries to find observations
containing PC mode XRT data and UVOT images obtained in the optical filters; if
successful, it uses these data to enhance the position. If no such observations
can be found, it reverts instead to using the UV filters, and the systematic
error is accordingly increased from 1.36\arcsec\ to 1.9\arcsec.

For GRBs, only observations which begin within 12 hours of the first one are
included; this is because the GRB is generally too faint to detect after this
point. Although this is not necessarily true for non GRB sources, this behaviour
is kept by default because using more observations increases the length of time
taken to produce the position, and our GRB experience shows that most positions
produced with this selection criteria are limited by systematics, not
statistics. This can be changed by the user, or the user can explicitly state which
observation(s) should be used. In the case of an object which was monitored for
some time in quiescence by \swift\ before undergoing some outburst, it is
particularly recommended that users specify the obsID to use, since  the
brighter data from the outburst would not be included by default but are more
likely to give a good position (unless the outburst pushes XRT into WT mode).

\subsection{Light curves}
\label{sec:ngrblc}

For most non-GRB objects the binning method used for GRBs -- defining bins by
the number of counts they contain -- is not ideal. We have thus produced a
version of the software which bins in a conventional way: the user specifies the
duration of the bins (the method used for binning GRB light curves is however
still available for non-GRB sources; alternatively users can choose to produce
one bin per snapshot, or per observation). After a gap in the data (for example,
between snapshots) a new bin begins at the start of the next GTI which is not
necessarily an integer number of bin widths after the last bin ended. There are
several caveats about this binning method:

1) The software uses Gaussian statistics to calculate the uncertainty after
background subtraction. If there are fewer than 15 counts in a bin, this may not
be accurate. It is the user's responsibility to choose a bin size which ensures
sufficient counts per bin. A warning is given if any bins contain fewer than
15 counts.

2) Sometimes the XRT enters WT mode for reasons other than the source being
bright. This can produce spurious light curve points with large error bars (both
are invalid), and disrupt the scaling of the plots. By default any WT mode bin
with fewer than 15 counts is assumed to be spurious and is not included in the
light curve produced. The web interface allows the user to change the minimum
number of WT counts necessary for a valid bin -- setting it to 0 will include
all WT data points.

3) The last bin in a snapshot may have a low fractional exposure, in which case
any statistical fluctuations in the data will be exaggerated. We recommend that
users check any such points and consider rejecting any point with a low
fractional exposure; an option to do this automatically is provided by the
interface.

When the fixed bin-width binning method is used, an OGIP compliant FITS file is
produced containing the light curve, in addition to the standard products
created for GRBs.

\section{Results}
\label{sec:res}

In Section~\ref{sec:prods} we presented details of how enhanced positions,
light curves and spectra are produced. These methods have been applied to 
every GRB detected by the XRT, and will run automatically on all new GRBs.
The results are posted online and are available via:

\begin{itemize}
\item{Index: http://www.swift.ac.uk/xrt\_products}
\item{Positions: http://www.swift.ac.uk/xrt\_positions}
\item{Light curves: http://www.swift.ac.uk/xrt\_curves}
\item{Spectra: http://www.swift.ac.uk/xrt\_spectra}
\end{itemize}

\noindent For SPER data the results are available via: http://www.swift.ac.uk/sper. Each
page contains detailed documentation, including a log of any changes made after
publication of this paper.  All of these pages are interlinked and an index of
these results, and those from the BAT, is available via the GCN ground analysis
web page at:  http://gcn.gsfc.nasa.gov/swift\_gnd\_ana.html.

In Tables~\ref{tab:enhpos}--\ref{tab:avespec} we list the enhanced
positions, best-fitting light curve parameters, and best-fitting
spectral fit results for all GRBs observed by the XRT up to GRB
080723B\footnote{In order to fit the table within an A4 page it is
necessary to tabulate the decay indices and break times separately. The
online data contain these together in a single file.}.  In
Table~\ref{tab:alphas} we also give the \swift\ target ID and the BAT
$T_{90}$ (from the Swift data
table\footnote{http://heasarc.gsfc.nasa.gov/docs/swift/archive/grb\_table.html/})
for reference. Although our automated processing only produces
time-averaged spectra, in this compilation of results it is interesting
to consider possible spectral evolution. So, for each  GRB with a break
in its light curve we extracted and fitted spectra for each light curve
segment (delimited by the breaks). These results are presented in
Tables~\ref{tab:resbeta} and~\ref{tab:resnh}\footnote{We have separated
$\beta$ and $N_{\rm H}$ for the paper, however these are in a single
table online.} The tables are also available online, through the Virtual
Observatory (ivo://uk.ac.le.star.swift/dsa\_grb\_aux/SwiftXRTGRBCat), and
via CDS (http://cdsweb.u-strasbg.fr/cgi-bin/qcat?J/MNRAS/).

%All of these tables are also
%available online  via CDS, and through the Virtual Observatory (\emph{\bf
%details will appear here!}). Note that errors quoted in all tables are at the
%90\%\ confidence interval.

\subsection{Validation of SPER results}
\label{sec:sperres}

The analysis of SPER data is intended to give users an indication of a GRB's
properties extremely rapidly; they are not intended for scientific analysis.
Since these data are in every case superceded by Malindi data, we do not list
the SPER results in this paper; however we demonstrate their veracity and the
limits thereof.

As with the enhanced Malindi positions, we determined the offset between the
enhanced SPER position and the UVOT position of every GRB with both of these
positions, and confirmed that they agreed 90\%\ of the time, i.e.\ the enhanced
SPER 90\%\ confidence error radius is correctly calibrated.

\begin{figure}
\begin{center}
\psfig{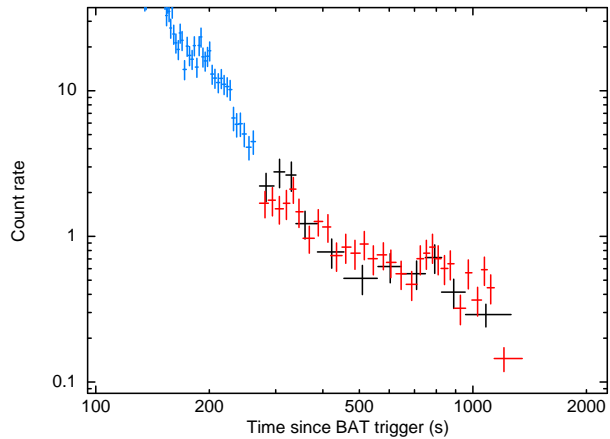}
\caption{An example SPER light curve (black) plotted with the Malindi-data light
curve (red -- PC and blue -- WT). The SPER data clearly give a good
representation of the true light curve.}
\label{fig:spermallc}
\end{center}
\end{figure}

We chose not to fit the SPER light curves as they typically have few bins.
Instead we compared SPER and Malindi curves by eye, an example is given in
Fig.~\ref{fig:spermallc}. We found good agreement between SPER and Malindi light
curves. 

To test the spectra, we created spectra from Malindi data covering the same time
region as the SPER data, and fitted them. We then compared the column density,
spectral index and observed flux between these fits and those from the SPER
spectra. The first two parameters were in good agreement, however the fluxes
only agreed within their 90\%\ errors 70\%\ of the time. This discrepancy
probably reflects the lack of GTI information for SPER data, and may also
suggest that using the covariance matrix from the fit to estimate the flux
errors (as we do) underestimates the errors.

\subsection{`Malindi' data}
\label{sec:malres}

The large volume of uniformly analysed Malindi data presented in
Tables~\ref{tab:alphas}--Table~\ref{tab:resnh} (i.e.\ temporal and spectral
analyses) allow us to consider the sample properties of the GRBs observed by
\swift's XRT.

We have created probability distribution functions (PDFs) for the temporal and
spectral indices and temporal break times. We prefer these to histograms as the
latter neglect uncertainties on the parameters; a PDF accounts for them and so
gives a better representation of the data. The sample PDFs  were obtained by
treating the PDF of any single parameter as two halves of normal distributions
with widths set by the measured uncertainties, each half normalised so as to
form a continuous function, as done by Starling \etal(2008), i.e.\

\begin{equation}
P(x|\bar{x},\sigma_1,\sigma_2)=\frac{\sqrt{2}}{\sqrt{\pi}\left(\sigma_1+\sigma_2\right)}
\left\{\begin{array}{ll}e^{-(x-\bar{x})^2/2\sigma_1^2} & (x\le\bar{x}) \\
e^{-(x-\bar{x})^2/2\sigma_2^2} & (x>\bar{x}) \end{array} \right. 
\,\end{equation}

\noindent where the 1-$\sigma$ errors on each parameter are taken as the
calculated 90\%\ confidence error divided by 1.6. We then created
overall PDFs of the temporal index ($\alpha$) and spectral index
($\beta$) by summing the PDFs of each individual $\alpha$ or $\beta$
parameter and dividing the merged PDF by the number of contributing
values. The peak and FWHM of the various PDFs are given in
Table~\ref{tab:pdfs}; these were calculated by fitting Gaussians to the
PDFs; note that many distributions are clearly more complex than a
simple Gaussian, in which case the values in Table~\ref{tab:pdfs} should
be taken as indicative, rather than precise.

\begin{figure}
\begin{center}
\psfig{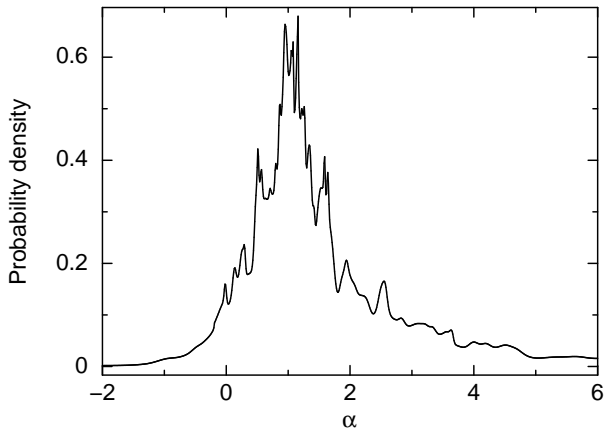}
\caption{Probability distribution function of the light curve power-law decay
indices tabulated in Table~\ref{tab:alphas}.}
\label{fig:alphaPDF}
\end{center}
\end{figure}

The $\alpha$ PDF is given in Fig.~\ref{fig:alphaPDF}, and shows a fairly tight
distribution of values. There are a total of 665  values suggesting that the
steep drops in probability around $\alpha=0.5,1.5$ are real. This is perhaps not
surprising, the `canonical' X-ray light curve (Nousek \etal2006; Zhang
\etal2006) contains four phases, and as discussed below there are several other
light curve morphologies observed, with one or more distinct phases. Each phase
has its own $\alpha$ distribution corresponding to the peaks in
Fig.~\ref{fig:alphaPDF}. See Section~\ref{sec:discuss} for more details.

\begin{figure}
\begin{center}
\psfig{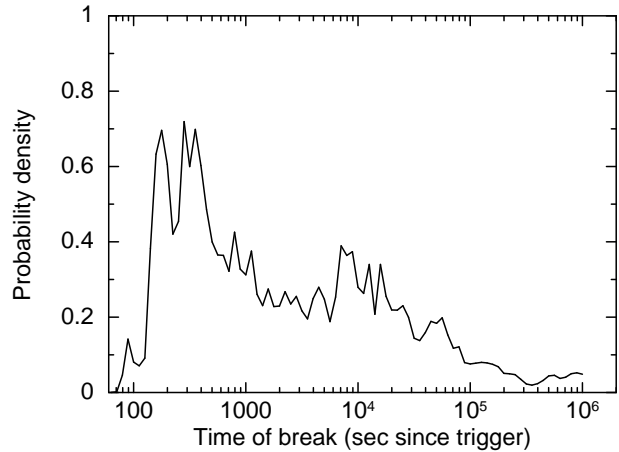}
\caption{Probability distribution function of the break times [probability per
unit log(time)] between the power-law decay segments, tabulated in Table~\ref{tab:breaks}.}
\label{fig:breakPDF}
\end{center}
\end{figure}

The PDF of observer-frame break times is given in Fig.~\ref{fig:breakPDF}; we
lack the redshift information necessary to translate to the rest frame for most
bursts. Since GRB light curves span many decades, the probability density is
defined here as probability per unit log(time), rather than per unit time. The
two peaks around \til1--300 s and 10$^4$ s arise from the `canonical' light curves and
reflect the  most common start and end times of the plateau phase, however there
is significant probability of a break at all times between \til100--10$^5$ s.
This is likely the result of two effects: the redshift distribution of GRBs, and
an intrinsic scatter of GRB light curve morphology. There are also selection
effects which may affect this distribution: to tightly constrain a break
requires good sampling of the decay on either side of the break, thus towards
the end of the light curve breaks are much harder to constrain and will have
broader PDFs, or not be seen at all (Curran \etal2008; Racusin \etal2009). Also, there is usually a
gap in \swift\ observations between \til2--4 ks after the trigger (while
\swift\ is the far side of the Earth from the GRB) making it harder to tightly
constrain breaks in this interval.

\begin{figure*}
\begin{center}
\psfig{file=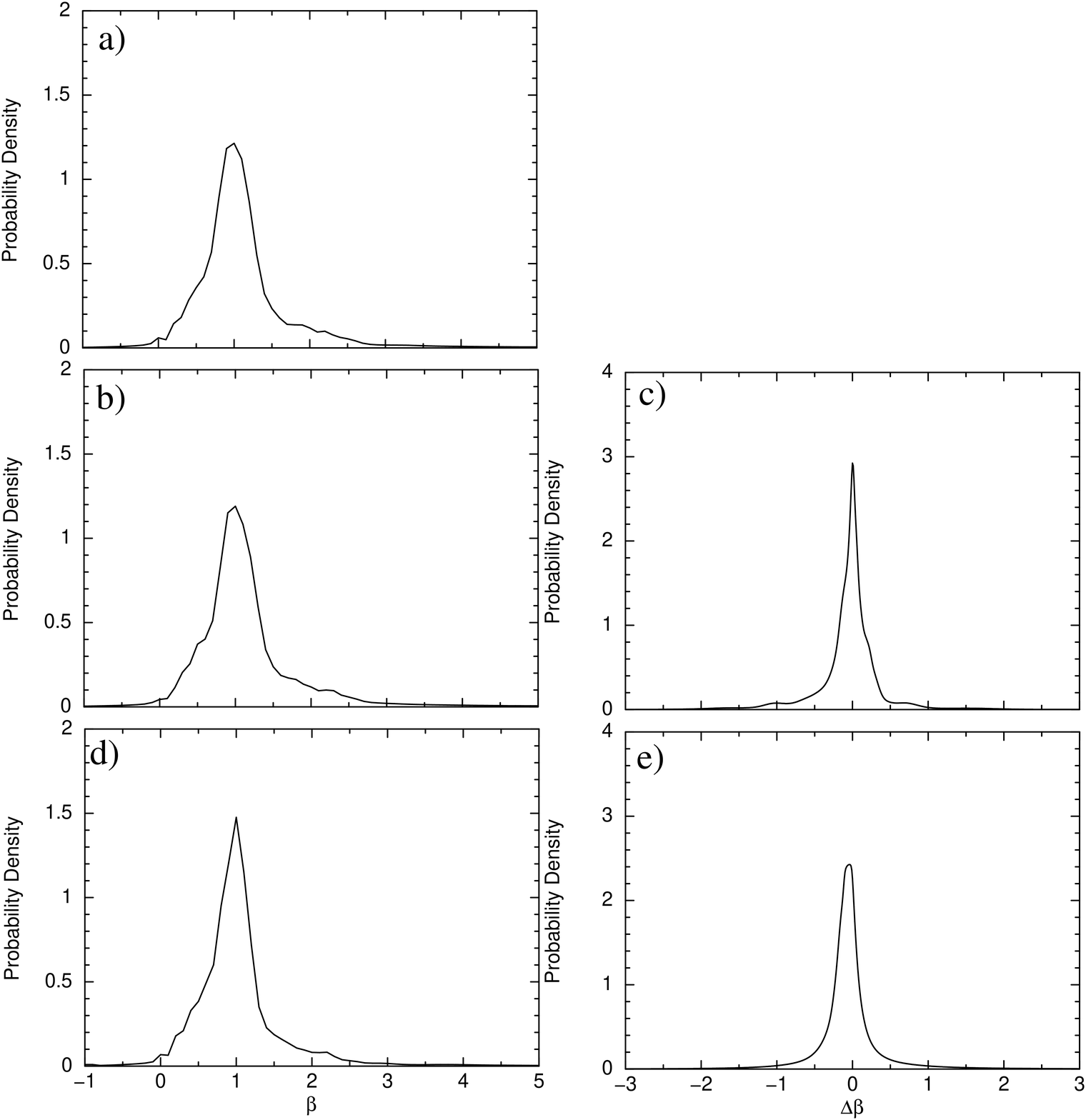,width=16cm}
\caption{Probability density functions (PDFs) of the spectral index ($\beta$)
from the time-averaged spectra. Panel a) shows the overall distribution. b)
shows this when times of flaring are excluded. Panel c) shows the PDF of
$\Delta\beta$ caused by removing flares. Panel d) shows the distribution when
bursts with unknown redshift are assumed to be at 2.23, rather than 0, and e)
shows the PDF of $\Delta\beta$ resulting from this change.}
\label{fig:gammaPDF}
\end{center}
\end{figure*}

Turning to the spectra, the PDF of the spectral index ($\beta$) is shown in
Fig.~\ref{fig:gammaPDF} (panel a). Here, and in Tables~\ref{tab:avespec}
and~\ref{tab:resbeta}, we give the spectral index $\beta$ (i.e. $F_\nu\propto
E^{-\beta}$). Some authors prefer the photon index $\Gamma$ (i.e.\ $N_E(E)\propto
E^{-\Gamma}$, this is the value used in the {\sc xspec} power-law model). These
are very simply linked: $\beta=\Gamma-1$.

In creating the spectra we made no attempt to exclude times of flares,
preferring to maximise the number of counts in the spectra. Previous studies
(e.g.\ Falcone \etal2007) and the X-ray hardness ratios on the XRT light curve
repository, show that flares tend to have harder spectra than the underlying
afterglow emission. Table~\ref{tab:flares} shows that 81 of the GRBs in our
sample contained `deviations' from power-law decays, many of which were flares.
To determine whether this has biased our results, we regenerated the spectra
excluding the times of any deviations from power-law behaviour, as identified in
the light curve fitting phase (Table~\ref{tab:flares}). In panel b) of
Fig.~\ref{fig:gammaPDF} we show the PDF of the spectral index from these data
(this includes all the GRBs with no deviations, as well as those where
deviations were removed), this is almost identical to that in panel a). In panel
c) we show the PDF of the change in $\beta$ caused by ignoring the times of
deviations (derived only from the 81 GRBs listed in Table~\ref{tab:flares}).
This shows the mean change to be 0$\pm0.2$; for comparison, the median
uncertainty in $\beta$ in Table~\ref{tab:avespec} is $\pm0.16$. We thus conclude
that the presence of flares has a negligible effect on our time-averaged
spectra.

Another factor which may affect the spectral index is the redshift ($z$) of the
burst; if the redshift used in the fit is incorrect and there is significant
absorption in the GRB host galaxy, the absorption will not be correctly modelled
-- this is most notable around the neutral oxygen edge at 0.525 keV (rest
frame). This in turn affects the spectral index. For example, for GRB 050904
($z=6.29$, Cusumano \etal2007) the fit reported in Table~\ref{tab:avespec} has
$\beta=0.927^{+0.048}_{-0.049}$. If we set the absorption above the Galactic
value to be at $z=0$ we find $\beta=0.964^{+0.062}_{-0.060}$.

In our default approach (Section~\ref{sec:spec}), unless the redshift of the GRB
has been spectroscopically determined, we use an unredshifted absorber to model
the excess absorption, as well as the Galactic component. Since this is clearly
incorrect, we tried refitting all these GRBs, using $z=2.23$\footnote{the mean
reported redshift for \swift-detected GRBs to date, calculated from
Table~\ref{tab:avespec}.} for the excess absorption if the redshift was not
known. In panel (d) of Fig.~\ref{fig:gammaPDF} we show the PDF of spectral
indexes where $z$ was taken from the literature, where available, and otherwise
set to 2.23. Panel (e) shows the PDF of the change in $\beta$ caused by assuming
$z=2.23$ instead of $z=0$ for those GRBs with no spectroscopic redshift.  The
latter is $0\pm0.2$, suggesting that it is acceptable to assume no redshift if
no spectroscopic determination has been made.

\begin{table}
\begin{center}
\begin{tabular}{ccc}
\hline
Parameter         &   Peak of PDF  &   FWHM   \\
\hline
$\alpha$     &  1.1           &   1.5    \\
$\beta$           &  0.98          &   0.66   \\
log(\nh) \cms\    &  21.3          &   1.2    \\
Counts-to-flux    &  3.8\tim{-11}  &  1.4\tim{-11} \\
factor (observed) & (erg \cms\ ct$^{-1}$) & (erg \cms\ ct$^{-1}$)\\

\hline
$\alpha^a_{\rm plat}$   & 0.32 &  0.79    \\
$\alpha^a_{\rm norm}$   & 1.2  &  0.5     \\
$\beta^a_{\rm plat}$    & 1.0  &  0.53    \\
$\beta^a_{\rm norm}$    & 1.1  &  0.65    \\
log($T_{rm plat,start}$) & 2.5 & 0.7 \\
log($T_{rm plat,end}$) & 4.0 & 1.2 \\
\hline
$\alpha^b_{\rm steep}$    & \til2.7 & \til2.1 \\
$\alpha^b_{\rm shallow}$     & 0.82    & 0.46\\
$\beta^b_{\rm steep}$     & \til1.1 & \til1.3 \\
$\beta^b_{\rm shallow}$   & 1.21    & 0.68\\
log($T_{\rm break}^b$) & 2.8 & 0.75 \\
\hline
$\alpha^c_{\rm steep}$    & \til1.2 & \til0.9 \\
$\alpha^c_{\rm shallow}$  & \til0.7 & \til0.8 \\
$\beta^c_{\rm steep}$        & 1.1     & 0.6   \\
$\beta^c_{\rm shallow}$   & 0.98    & 0.38 \\
log($T_{\rm break}^b$) & \til4.0 & \til2.8 \\
\hline
$\alpha^d_{\rm steep}$       & 1.1     & 0.7  \\
$\beta^d_{\rm steep}$        & 1.1     & 0.6  \\
\hline
\end{tabular}
\caption{The peak and FWHM of the PDFs plotted in
this paper, where the PDFs have been approximated
as Gaussians. $^{a-d}$ refers to
the different classes of GRB illustrated in Fig.~\ref{fig:schematic},
e.g.  $\alpha^b_{\rm steep}$ refers to the decay index of the
steep-decay phase of those light curves which begin steep and then
flatten.
\newline Note: the underlying distributions are not always Gaussian, or
are poorly sampled so the values herein are indicative.
}
\label{tab:pdfs}
\end{center}
\end{table}

\begin{figure*}
\begin{center}
\psfig{file=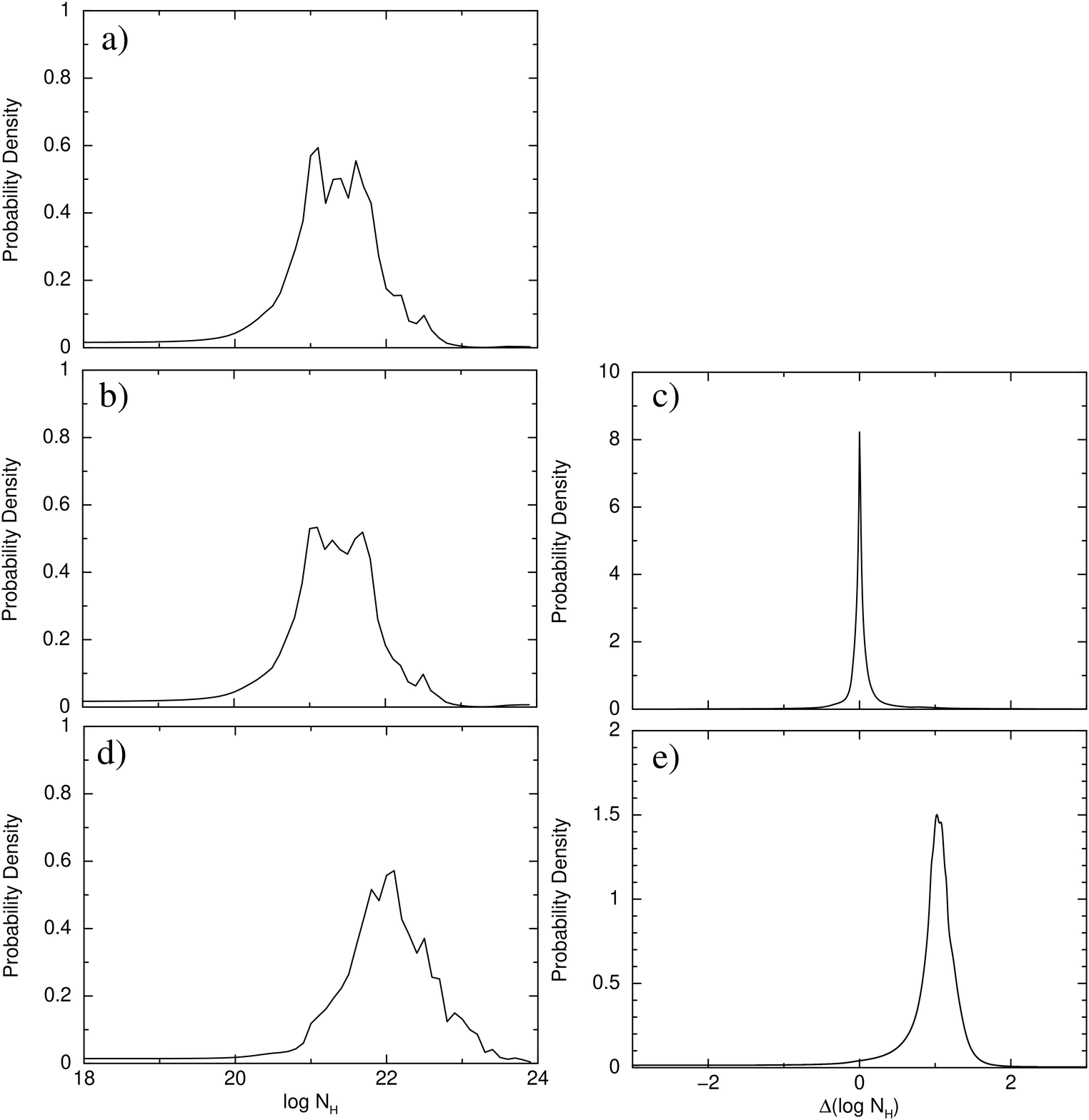,width=16cm}
\caption{Probability density functions (PDFs) of the `excess' column density,
i.e.\ that above the Galactic value, from the time-average spectra.
Panel a) shows the overall distribution. b) shows this when times of flaring are
excluded. Panel c) shows the PDF of $\Delta \log \nh$ caused by removing flares.
Panel d) shows the distribution when bursts with unknown redshift are assumed to
be at 2.23, rather than 0, and e) shows the PDF of $\Delta \log \nh$ resulting from
this change.}
\label{fig:nhPDF}
\end{center}
\end{figure*}

In Fig.~\ref{fig:nhPDF} we present the PDF of the `excess' absorption: that
is, the value of the second absorption component in the fit (the first being
frozen at the Galactic value from Kalberla \etal2005, see Section~\ref{sec:spec}
above). The panels are as in Fig.~\ref{fig:gammaPDF}.

\subsubsection{Time-resolved analysis}
\label{sec:restimeres}

It is interesting to consider how the sample of \swift\ GRBs presented here
compare to theoretical predictions. In this paper we limit ourselves to
comparison with the fireball model (e.g.\ Sari \etal1998), introduced in
Section~\ref{sec:fireball}. We do this using the `closure relationships' which
related the temporal index ($\alpha$) and the spectral energy index ($\beta$),
see, e.g.\ Zhang \etal(2006) for a table of such relationships. These are
simplifications of the complete model; in particular they assume that the
electron distribution $p\til2.2$, and that the microphysical parameters such as
the proportion of blastwave energy stored in magnetic fields are not evolving
through the outburst.

\begin{figure*}
\begin{center}
\psfig{file=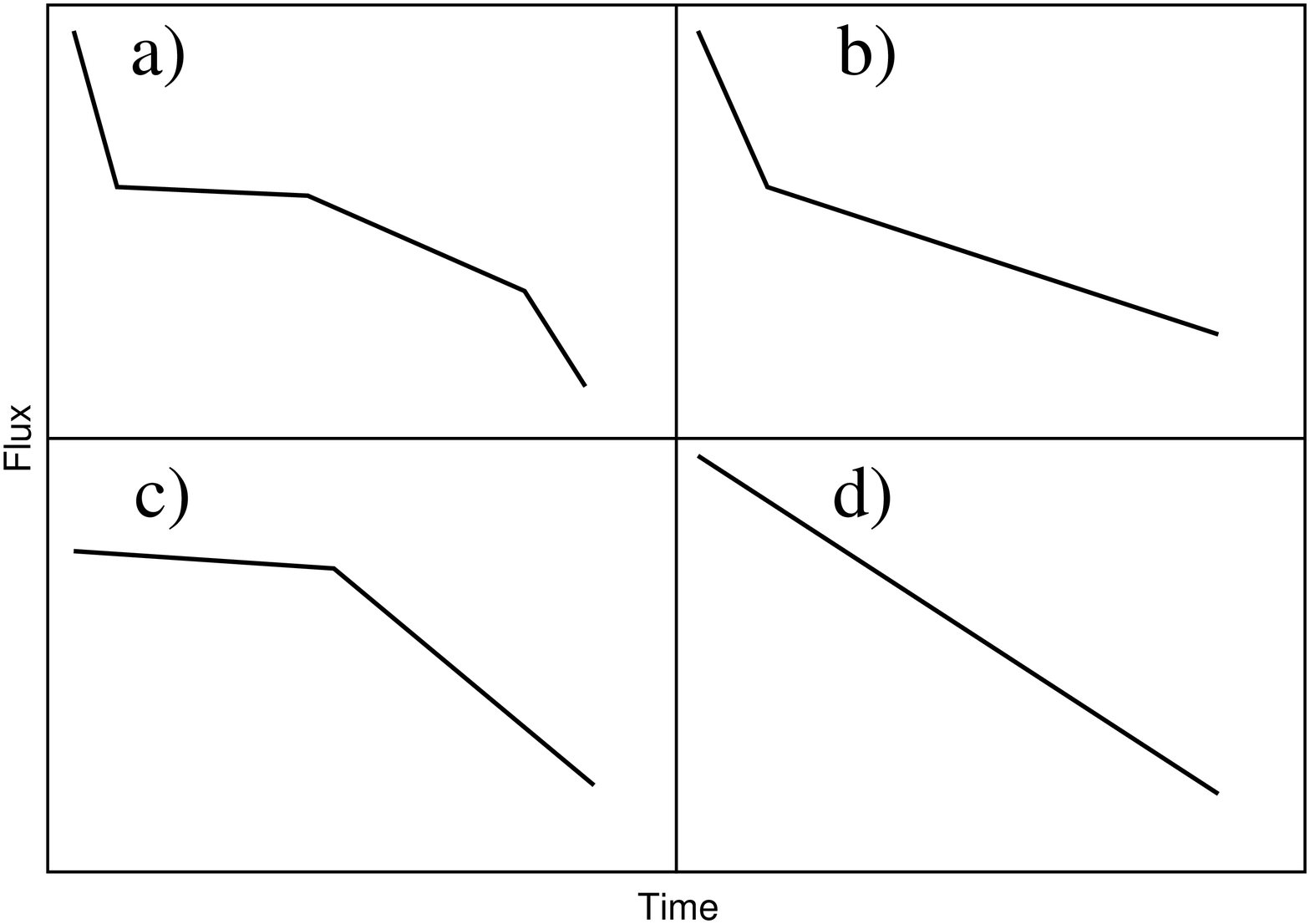,width=16cm}
\caption{Schematic diagrams of the different light curve morphologies seen,
excluding the `oddballs'. Panel a) shows the so-called `canonical' light curves.
Panels b)--c) are those with one break, either flattening (b) or steepening (c).
Panel d) are those with no breaks.}
\label{fig:schematic}
\end{center}
\end{figure*}

Since the `canonical' afterglow (Nousek \etal2006; Zhang \etal2006) contains 4
distinct phases -- high-latitude emission (which is not afterglow emission), a
plateau, a `normal' decay phase, and post-jet-break decay -- it is not sensible
to compare all of the $(\alpha,\beta)$ pairs we have derived \emph{en masse}.
Instead we classified each light curve segment to study the groups separately.
To achieve this, we first classified each light curve either as `no break', `one
break', `canonical' or `oddball'. The first two are self explanatory. For the
latter two: any light curve with at least two breaks was considered canonical if
it contained one shallowing break, with $\Delta\alpha\leq-0.5$, and a later
steepening break, with $\Delta\alpha\ge0.5$, and an oddball otherwise [note that
the canonical light curve, as defined by Nousek \etal(2006) does not contain
this quantitative definition; we provide it for homogeneity]. We manually
checked these classifications and reclassified 7 light curves from canonical to
`oddball' (for example GRB 060202 shows a steep-shallow-steep-shallow behaviour
which is not canonical, but meets the criteria defined above).  The list of
light curves in each class is given in Table~\ref{tab:types}, and schematic
diagrams of all classes except `oddball' (which comprises a range of
morphologies and cannot be shown schematically) are given in
Fig.~\ref{fig:schematic}. The two types of singly-broken decay morphologies will
be referred to as `type b' and `type c' morphologies hereafter, for brevity.

For each canonical light curve we defined any segment with a positive $\alpha$
(i.e.\ decaying) before the break with $\Delta\alpha\leq-0.5$ as belonging to
the steep decay phase. Segments after this, but before the  break with
$\Delta\alpha>0.5$ were identified as the plateau phase. The next segment was
identified as the `normal' decay phase and any subsequent decay segments were
assumed to be post-jet-break decays. These classifications were again checked by
eye, and a small number of segments reclassified accordingly.

\begin{figure*}
\begin{center}
\psfig{file=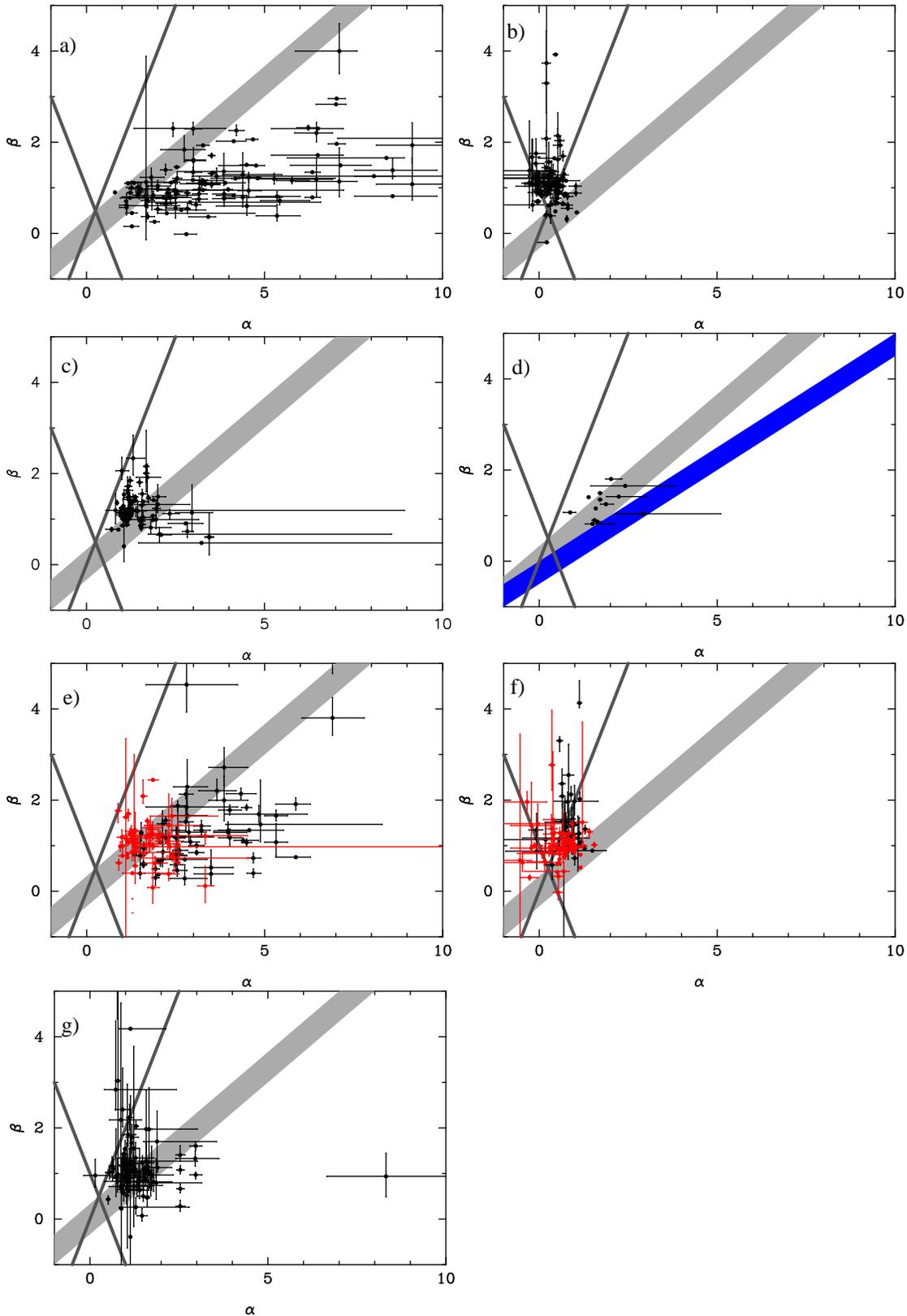,height=21.5cm}
\caption{Spectral indices ($\beta$) vs temporal indices ($\alpha$) for different
light curve phases; see Fig.~\ref{fig:schematic}. Panels a)--d) are for the
`canonical' light curves and show the values from the steep decay, plateau,
`normal' and `late break' phases. Panels e)--f) show the values from those light
curves showing a single break; the steeper of the segments are plotted in e) and
the shallower in f). The black and red points indicate type b and c light curves
respectively (see Fig.~\ref{fig:schematic}). Panel g) show the values for those
light curves which do not contain a break. The grey bands mark the areas
permitted by standard afterglow closure relationships; the narrow grey lines are
for the fast-cooling regime. The blue band in panel d) marks the range permitted
by post-jet-break closure relationships.}
\label{fig:closure}
\end{center}
\end{figure*}

\begin{figure*}
\begin{center}
\psfig{file=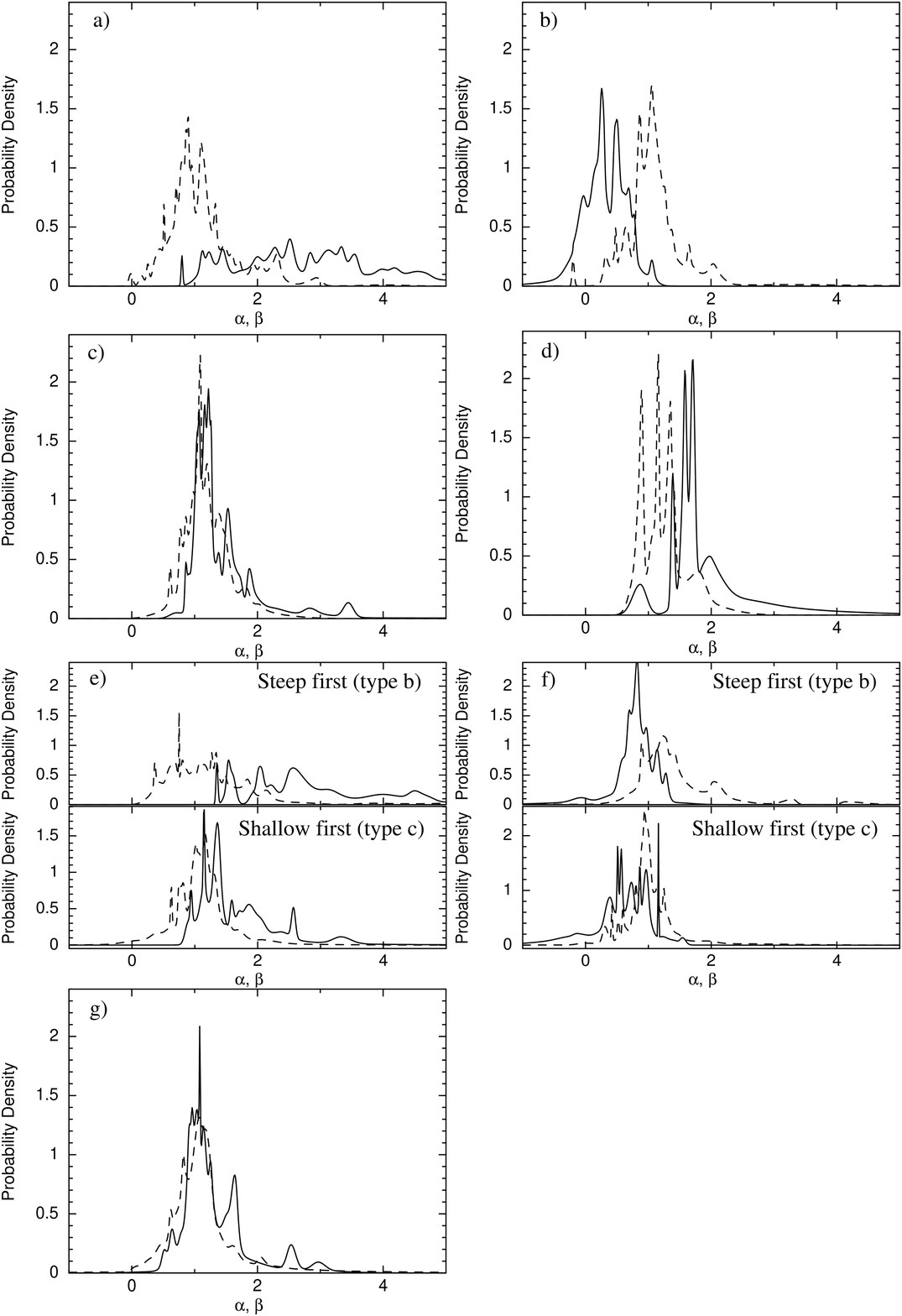,height=22cm}
\caption{PDFs of the temporal (solid lines) and spectral (dashed lines) indices
($\alpha$ and $\beta$ respectively) for the different light curve phases. 
Panels a)--d) are for the `canonical' light curves and show
the values from the steep decay, plateau, `normal' and `late break' phases.
Panels e)--f) show the values from those light curves showing a single break;
the steeper of the segments are plotted in e) and the shallower in f). For
clarity, these plots have been split into two panes to separate the
type b and c cases. Panel g) show the values for those
light curves which do not contain a break}
\label{fig:subpdfs}
\end{center}
\end{figure*}

In Fig.~\ref{fig:closure} we plot $(\alpha,\beta)$ from Tables~\ref{tab:alphas}
and~\ref{tab:resbeta} for each of these 4 segments of the canonical afterglow
(panels a--d;  PDFs of these are given in Fig.~\ref{fig:subpdfs}) We also
show the regions covered by the standard afterglow closure relationships (from
Zhang \&\ M\'esz\'aros 2004); the thick grey band shows the range allowed by the
pre-jet break emission in a slow cooling regime (i.e.\ the synchrotron peak
frequency, $\nu_m$ is below the cooling frequency $\nu_c$). The two dark grey
lines map the closure relationships for the fast cooling regime ($\nu_m>\nu_c$).
In panel (d) we also show the region covered by the post-jet break closure
relationships (blue band). The points in panels a) and b) appear uncorrelated
with the closure relationships, however this is unsurprising. In the standard
interpretation, the steep decay phase (a) is not afterglow emission, and in the
plateau phase (b) energy is being injected into the afterglow. What is perhaps
surprising is that many of the `normal decay' phase points (c) do not lie within
the range predicted by afterglow theory. The post-jet-break points (d) are the
only ones which agree well with afterglow theory, although they agree much
better with the pre-jet break relationships than the post-break ones. We
consider these facts in more detail below (Section~\ref{sec:interpret}).

For the bursts showing only one break, we show the $(\alpha,\beta)$ values of
the steeper decay in panel (e) of Fig.~\ref{fig:closure} and the shallower decay
in panel (f). The black points are type b curves and the red ones type c (c.f.\
Fig.~\ref{fig:schematic}).Panel (g)
shows the bursts with no breaks in their X-ray light curves. Overall,
Fig.~\ref{fig:closure} shows similar results to those in Butler \&\ Kocevski
(2007a, fig 1), however they had a smaller sample and divided the light curves
into segments based on a uniform time-slice rather than individual fits, thus our
results are not directly comparable.

\begin{figure}
\begin{center}
\psfig{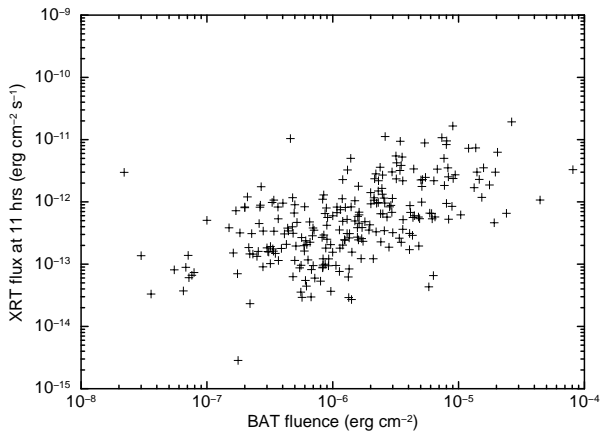}
\caption{The X-ray flux at 11 hours post-trigger plotted against the BAT
fluence. The correlation reported by Gehrels \etal(2008) is still present in our
larger dataset. Note that Gehrels differentiated between long and short bursts (the
correlation being more obvious for the former) which we have not done.}
\label{fig:lx}
\end{center}
\end{figure}

\begin{figure}
\begin{center}
\psfig{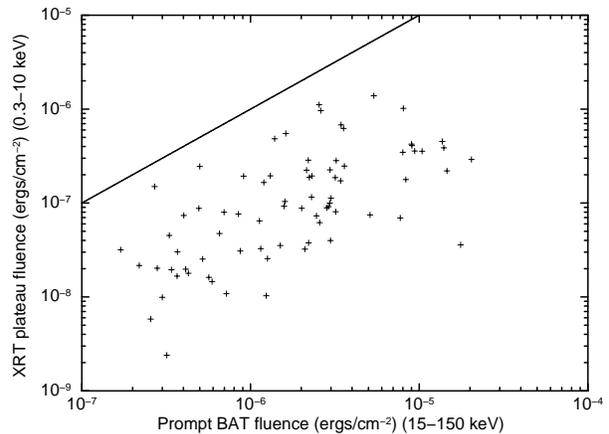}
\caption{The fluence of the X-ray plateau phase plotted against the BAT fluence
of the prompt emission, for the canonical light curves. The line shows where the
two quantities are equal. This plot is analogous to
fig.~6 of Willingale \etal(2006) but with a larger sample, and confirms their
findings.}
\label{fig:fluence}
\end{center}
\end{figure}

\begin{figure*}
\begin{center}
\psfig{file=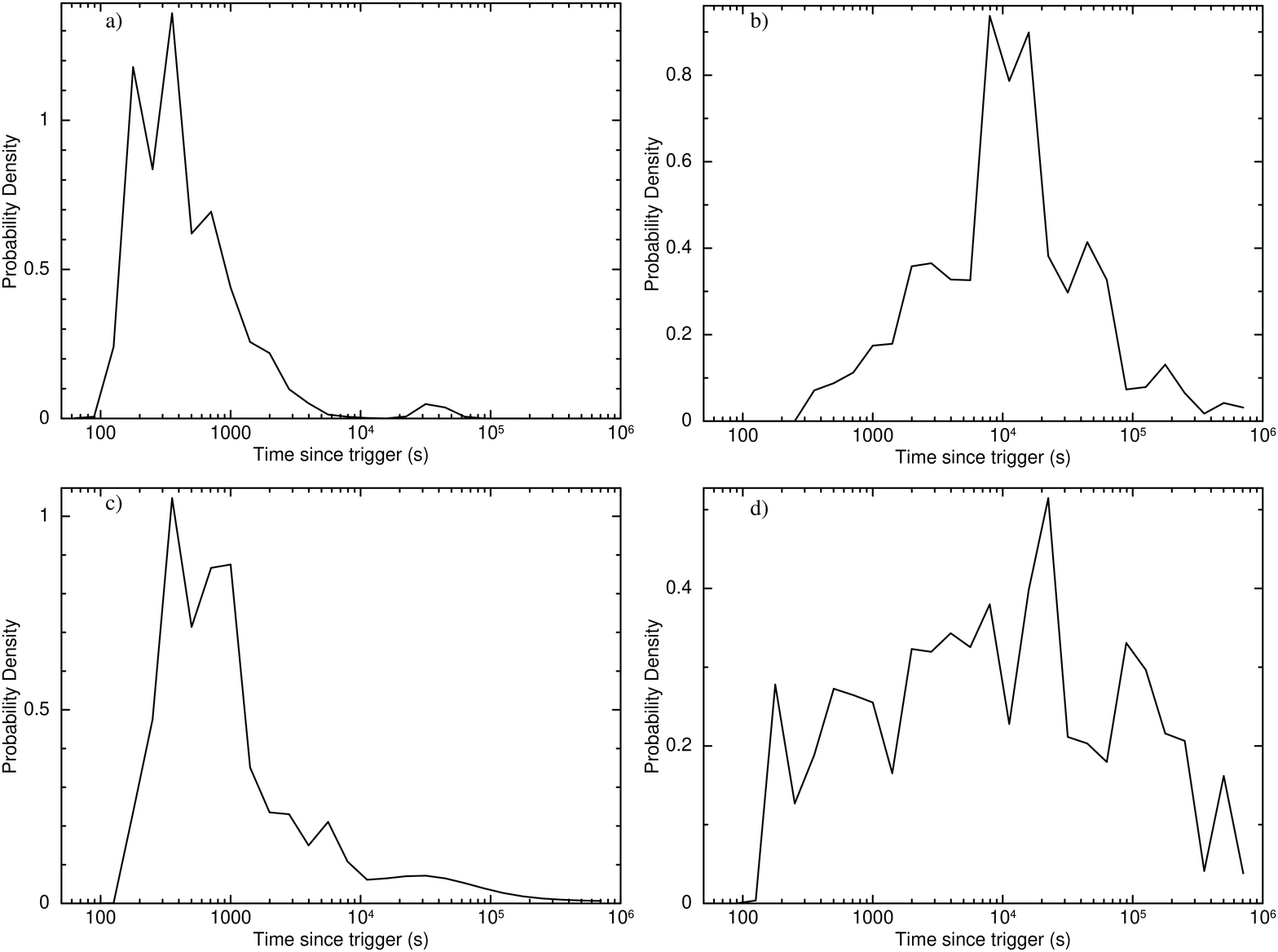,width=16cm}
\caption{PDFs of the break times for the canonical light curve and curves with one break.
Panels a) and b) give the start and end times respectively of the plateau phase of canonical light curve.
Panel c) shows the break times for light curves which show a single, flattening break (type b) and
panel d) gives the break times for those which show a single, steepening break (type c).}
\label{fig:breakPDFs}
\end{center}
\end{figure*}

\subsection{Other sample statistics}
\label{sec:randomstats}

In the rest of this paper we will concentrate on the collection of light
curves presented in this catalogue. These data can also be combined with
other datasets such as the BAT or UVOT catalogues (Sakamoto \etal2008;
Roming \etal2009) or the online \swift\ data
table\footnote{http://heasarc.gsfc.nasa.gov/docs/swift/results/}.

For example, Gehrels \etal(2008) compared $E_{\rm iso}$ with the X-ray
luminosity 11 hours after the trigger and found a correlation with a slope of
\til1. They thus suggested that the radiative efficiency of the blastwave is
similar from GRB to GRB.  In Fig.~\ref{fig:lx} we show the BAT fluence plotted
against the X-ray flux at 11 hours (i.e.\ we have made no correction for
distance, since only \til30\% of the GRBs in our catalogue have known redshift)
using the large sample of GRBs in this paper; the trend can still be seen.

As another example, Willingale \etal(2007) compared the fluence of their two
emission components (prompt and afterglow), found a weak correlation and showed
that that the afterglow fluence never exceeds the prompt fluence. We have
performed an analogous analysis, comparing the 15--150 keV BAT fluence from the
\swift\ data table with the fluence of the plateau phase in the canonical light
curves (Fig.~\ref{fig:fluence}). Note that Willingale \etal(2007) determined the
fluence over a wide energy band wheras we have only considered the 2 distinct
bands covered by the data (i.e.\ 0.3--10 keV for XRT and 15--150 keV for BAT).
Our results are however consistent with those of Willingale \etal(2007).
Considering the plateau further if it is caused by energy injection (see below)
one may na\"ively expect to see some relationship between $T_{90}$ and the
plateau duration, e.g.\ perhaps longer lived bursts also inject energy  for
longer. Combining our data with the \swift\ data table however reveals no
correlation between these two parameters.

These are a few examples of the large-scale studies which our dataset enables;
to aid in such studies all of our light curves, positions and spectra are online
and the tabulated data in this paper are available in machine readable format;
details were given earlier in this section.

\section{A canonical light curve?}
\label{sec:discuss}

 For the rest of this paper we consider the sample of GRB light curves, and
specifically, the range of morphologies found, as demonstrated in
Fig.~\ref{fig:schematic}. Nousek \etal(2006), Zhang \etal(2006) and Panaitescu
\etal(2006) proposed that there is a `canonical' X-ray GRB  light curve,
consisting of 4 power-law phases: a steep initial decay, a shallow plateau, and
then a `normal' decay which is steeper than the plateau, but not as steep as the
first segment. There may also be a fourth segment, post-jet break decay. Panels
c) and d) of Fig.~\ref{fig:autofit} show light curves which conform to this
behaviour. Nousek \etal noted that we do not see this exact behaviour in all GRB
afterglows, and suggested that this is simply due to limited temporal coverage.
O'Brien \etal(2006) meanwhile showed that, for GRBs observed by \swift, the
prompt emission seen by the BAT transitions smoothly into the emission seen by
the XRT.

Willingale \etal(2007) interpreted the observed X-ray emission as the
combination of two components, each following a simple exponential-to-power-law
form. A late-time break in the power-law may be added occasionally as well.
Physically, the two exponential-to-power law components were identified with the
prompt GRB emission from internal shocks in the ejecta, and afterglow emission
from an external shock in the circumburst medium. Under this model, not all GRBs
exhibit all segments of the `canonical' curve. For example, the afterglow
component can be sufficiently weak compared to the prompt component that it is
never seen, alternatively it can dominate from an early time.

In this paper we have presented a sample of GRBs much bigger than those used by
Nousek \etal or Willingale \etal (who used 27 and 107 bursts respectively), and
can thus consider the possibility of a unified afterglow model with more
confidence.

We defined a subset of the bursts presented in this paper, comprising only those
for which we can reasonably expect to have seen the three phases of the
canonical light curve. By inspecting the break times (Table~\ref{tab:breaks}) of
the canonical light curves (Table~\ref{tab:types}) we defined such bursts as
having XRT data beginning at $T\le T_0+200$ s and extending to at $T\ge T_0+50000$
s. We also specified that the light curve should contain at least 20 bins. This
gave a sample of 162 GRBs, which are shown in bold type in
Table~\ref{tab:types} (Note that previous studies, e.g.\ O'Brien \etal(2006); 
Willingale \etal(2007), did not create such subsets; their samples are analogous
to our complete sample, not this subset). Of these 162 bursts:
\begin{itemize}
\item{7 (4 \%) have no breaks.}
%\item{49 (30 \%) have one break (25 flatten, 24 steepen).}
\item{49 (30 \%) have one break (25 type b, 24 type c).}
\item{68 (42 \%) are canonical.}
\item{38 (24 \%) are oddballs.}
\end{itemize}

This immediately shows that the `canonical' light curve, while the most
common morphology actually occurs in less than half the GRBs in which it
would be identifiable. Whether or not the underlying afterglow behaviour
follows a single behaviour is readily testable with our  dataset and we
discuss the different light curve morphologies in this context below. To
aid this discussion, we show in Fig.~\ref{fig:subpdfs} the $\alpha$ and
$\beta$ PDFs for the various light curve phases, (the panels are as in
Fig.~\ref{fig:closure}) and in Fig.~\ref{fig:breakPDFs} PDFs of the
break times for the different light curve morphologies. These figures
were built using all 318 GRBs presented in this paper, not just the
subset defined above. Immediately, we see from this that the $\beta$
values are roughly the same in each light curve segment, although the
steep-decay phase of the canonical curves and the steep portion of the
one break, steep-to-shallow light curves have a wider range of values
than the other phases (a similar conclusion was drawn by Butler \&\
Kocevski 2007a). Note that we do not consider the `oddball' bursts here
since these need to be studied individually, whereas we are interested
in the bulk properties of afterglows. Schematics of the morphologies
discussed below were given in Fig.~\ref{fig:schematic}.

\subsection{Light curves with no breaks}
\label{sec:nobreak}

Although we have only seven GRBs with no breaks (Fig.~\ref{fig:schematic},
panel d) in our subsample, the distribution of $\alpha$ values from these GRBs
is inconsistent with the distribution from any phase of the canonical GRBs in
Table~\ref{tab:types} except for the normal decay phase. Even this consistency
is low; the K-S test gives a 1.9\%\  probability that the $\alpha$ values of
these 8 GRBs were drawn from the same sample as the normal decay phase of the
canonical GRBs (note that the K-S test is not necessarily believable with such a
low number of values). This suggests that the no-break GRBs are consistent with
the Willingale \etal(2007) model provided we are seeing only the power-law phase
of the afterglow component, however the lack of a steep decay phase  means that
either the prompt emission must decay very rapidly or the afterglow must be
bright enough to dominate from a very early time. Further, the lack of a plateau
phase implies that energy injection does not dominate at any time; the GRBs in
this subset (panel g of Fig.~\ref{fig:closure}) show reasonable agreement with
standard afterglow theory, supporting this idea. The outlier in that panel with
$\alpha$\til8 is GRB 051221B, and is a candidate `naked' GRB (Willingale
\etal2007).

\subsection{Light curves with one break: type b (shallowing decays)}
\label{sec:steeper}

Under the Nousek/Willingale models, type b light curves should
correspond to the first two segments of a `canonical' light curve, with
the plateau phase on-going when observations cease. In terms of
Figs~\ref{fig:closure}--\ref{fig:subpdfs}, this means that the black
points (upper pane in Fig.~\ref{fig:subpdfs}) in panel e) should come
from the same parent population as those in panel a), and those in panel
f) from the same population as those in panel b). By eye, the first of
these statements seems believable, and a K-S test gives a 17\%\
probability that the $\alpha$ values of the two samples came from the
same parent population. However, the $\alpha$ values of the shallow
decay in these light curves and the plateau phase of the canonical ones
are completely different. From Fig.~\ref{fig:subpdfs} one can see that
the distribution in the upper pane of panel f) ($\bar\alpha=0.0.85$)
lies towards significantly higher $\alpha$ than those in panel b)
($\bar\alpha=0.34$); a K-S test gives a $<0.1\%$ chance that the two
come from the same parent population. Further, although the distribution
of plateau start times (Fig.~\ref{fig:breakPDFs}, panel a) looks similar
to that of the type b break times (Fig.~\ref{fig:breakPDFs}, panel c),
the latter are shifted towards later times; a K-S test gives a 0.3\%\
chance that these share a common population.

It is still possible to reconcile the bursts with a single, steep to shallow
break to the same behaviour as the canonical bursts, if the shallow phase is
similar to the plateau phase, but the energy injection in these bursts is longer
lived and at a lower rate than in the `canonical' bursts. A rigorous
investigation of this is beyond the scope of this paper, and will be tackled in
a future publication. We do note however that if this is true, energy injection
must continue at least to the end of the \swift\ observations, which in all but
3 of these cases, is more than a day (often many days) post-trigger. Producing
such long lived energy injection at the necessary level, from the standard GRB
progenitor models is difficult, however X-ray flares have been seen $>1$ day
after the trigger (e.g.\ GRB 050502B, Falcone \etal 2006; GRB 080810, Page \etal
in prep.; see Curran \etal2008 for a discussion of late-time X-ray flares),
implying that the central engine can still affect the afterglow on these
timescales.

\subsection{Light curves with one break: type c (steepening decays)}
\label{sec:steeper}

Compared to the canonical light curve, type c light curves in could correspond
to the normal and post-jet break phases of a GRB light curve.  From an
$(\alpha,\beta)$ point of view this is acceptable; K-S tests show $>1\%$
probability that the red points (lower pane) in panel e) of
Figs.~\ref{fig:closure}--\ref{fig:subpdfs} come from the same population as
those in panel c), and those in panel f) come from the same population as panel
d). However, for this to be the case 7/25 (=28\%) of the `jet breaks' would have
to occur within 1000 s of the GRB trigger, suggesting an extremely confined jet.
Alternatively, the `jet breaks' in the canonical light curves may not be jet
breaks at all; this is suggested by panel d) of Fig.~\ref{fig:closure} and we
discuss this further in Section~\ref{sec:interpret}.

Instead of the above, the two phases of these GRBs could be identified
with the plateau and normal decay phases of the canonical light curve.
The distribution of plateau end times (Fig.~\ref{fig:breakPDFs}, panel
b) is similar to the (poorly sampled) distribution of type-c break times
(Fig.~\ref{fig:breakPDFs}, panel d), and a K-S test gives a 40\%\
probability that these represent the same population of times. However,
a K-S test between the decay slopes of the shallow part of the type c
light curves and the plateaux of canonical bursts gives a $<0.1\%$
probability that these come from the same population. This does not
definitively rule out this interpretation: if the afterglow dominates
the X-ray light curve before the prompt component decays, the effect of
energy injection may be less than in a canonical GRB, giving a steeper
shallow decay slope, as seen in our data. To investigate, we obtained
the BAT fluence from the Swift data table\footnote{via
http://heasarc.gsfc.nasa.gov/docs/swift/results/}  for all of the
canonical GRBs in Table~\ref{tab:types}, and for the one break,
shallow-to-steep GRBs from our subsample. If the latter bursts have
systematically lower fluence than the canonical GRBs, the above
explanation holds. No such trend is seen however.  %This test is not
strong enough to rule out %the explanation; to do so requires building
and fitting flux-unit light curves %from BAT and XRT for our sample and
the canonical GRBs, and will be done in a %future publication.

\section{Understanding the X-ray afterglow}
\label{sec:interpret}

We have shown above that the different morphologies of GRB light curves are
consistent with the two component model of Willingale \etal(2007); implying a
consistent underlying behaviour (if not a canonical shape). We now consider what
physical processes drive each of the phases obtainable from such a light curve.
The large, homogeneously-generated data set in this paper is an ideal test bed
for this. The usual explanation of the phases (e.g.\ Nousek \etal2006; Zhang
\etal2006, O'Brien \etal2006) is as follows: 

\begin{itemize}
\item{Steep decay -- high latitude prompt emission (internal shocks).}
\item{Plateau -- emission from a collimated external forward shock (afterglow) which is
undergoing energy injection. The edge of the jet is not visible to the observer.}
\item{Normal decay -- emission from a collimated external forward shock with no energy
injection. The edge of the jet is not visible to the observer.}
\item{Post jet-break -- emission from a collimated external forward shock with
no energy injection. The edge of the jet is visible to the observer.}
\end{itemize}

To compare our data with theoretical predictions for the steep decay phase
requires modelling of the BAT data, since $\alpha$ in this regime is sensitive
to $T_0$, which should be taken as the start time of the final pulse. This is
beyond the scope of our XRT-data paper, however many other authors have
confirmed that the steep decay phase is consistent with the expectation for high
latitude emission (e.g.\  Tagliaferri \etal2005; Barthelmy \etal2005b;
O'Brien \etal2006; Goad \etal2006; Liang \etal2006; Willingale \etal2007; Butler
\&\ Kocevski 2007b).

The plateau phase likewise is in good agreement with the above model. Zhang
\etal(2006) give the closure relationships for the energy injection scenario,
assuming the luminosity of the injecting source $L\propto t^{-q}$, where
$q\le1$. The lower edge of the solid grey band plotted in Fig.~\ref{fig:closure}
corresponds to the most tolerant $q=1$ limit from such relationships (as does
the blue band in panel d for the post-jet break, energy injection relationships
taken from Panaitescu \etal2006). Points lying above this line are
consistent with energy injection afterglow theory. 

The normal decay phase is not in good agreement with the interpretation above.
Panel c) of Fig.~\ref{fig:closure} shows many of the points to be inconsistent
with afterglow theory without energy injection. This disagreement continues into
the post jet-break phase (panel d), where the majority of points are consistent
with the standard pre-jet break models, but very few are consistent with post
jet-break theory. This suggests that the interpretation of the X-ray light curve
as given above is incorrect unless energy injection continues for some time
after the burst. (Note that we have not given an exhaustive study of jet breaks
as we include only breaks which occur after the first three `canonical' phases.
For a targeted study of potential jet breaks in any light curve morphology, see
Racusin \etal2009). 

Other models have been proposed instead to explain the observed X-ray light
curves. For example, Ghisellini \etal(2007) suggest that the long-term X-ray
emission is actually `late prompt' emission, from internal shocks with lower
bulk Lorentz factors than in the initial case. They use this to model X-ray and
optical light curves with some success (Ghisellini \etal2008). However in such a
model we may expect to see spectral evolution in the late-prompt emission
analogous to that seen during the prompt emission; as Table~\ref{tab:resbeta},
Fig.~\ref{fig:subpdfs}, and the hardness ratios on the XRT light curve
repository (Evans \etal2007) show, there is very little spectral evolution seen
in XRT data after the first few hundred seconds post-trigger (O'Brien \etal in
preparation). Note that the Ghisellini model implicitly assumes that the
late-prompt component does not spectrally evolve, so does fit the observed
hardness ratios, however why it does not evolve is not clear. The
dust-scattering model of Shao \&\ Dai (2007) suffers from the same problem (Shen
\etal2008). de~Pasquale \etal(2008) recently suggested that the end of the X-ray
plateau could signify a jet break, where energy injection is ongoing, subsequent
`jet breaks' would then signify the end of energy injection. However, this is
not consistent with the points in panel d) of Fig.~\ref{fig:closure} -- data
taken after the end of the plateau and a subsequent break -- which are generally
inconsistent with post jet-break models with no energy injection.

The dataset presented in this paper represents the best diagnostic tool for
afterglow models currently available, and can be used to place specific
constraints on any given model for the X-ray emission. For example, considering
the external forward shock model, Fig.~\ref{fig:closure} tells us that:

1) During the plateau phase, energy must be injected into the shock.

2) The so-called `post jet-break' phase in the `canonical' light curve is in
fact better explained as occurring before the jet break but after the cessation of
energy injection than by the standard interpretation of arising after the jet-break
and cessation of energy injection.

3) Some mechanism must cause a steepening of the light curve, independent of
energy injection. It must not invoke any spectral change.

The latter point arises because the break seen between the plateau and normal
phases cannot always be caused by the cessation of energy injection: too many
points in panel c) lie above the grey band hence must be undergoing energy
injection. 

There is a reasonable number of bursts whose normal decay phase is
consistent with a standard forward shock with no energy injection (i.e.\ points
in Fig.~\ref{fig:closure} panel c which lie within the grey band) as well as
many which do require energy injection during this phase. Thus in the
description above it must be possible for the unknown-origin break (in point 3
above) to occur before, or after the cessation of energy injection. Before this
break while energy injection is ongoing, a GRB lies on panel b) of
Fig.~\ref{fig:closure}, after the break and once energy injection has ceased
it lies on panel d). Whether the cessation of energy injection or the
unknown-origin break occurs first would then determine whether the GRB lies in
or above the grey band during its time on panel c).

There is a significant number of bursts lying above the grey band permitted by
the closure relationships in Panel c) of Fig.~\ref{fig:closure}, which do not
show a subsequent break. Similarly, the black points in Panel f) represent the
last observed state for many GRBs. This implies that for external, forward shock
model of the afterglow, significant energy injection must last for days, if not
weeks after the trigger. Ghisellini \etal(2008) suggest that this is possible,
however it is not clear that their mechanism can produce sufficient levels of
energy injection to sustain the shallow decay. Nonetheless, if energy injection
from lasting central engine activity (or slow-moving shells ejected at the time
of the burst) is responsible for the shallower-than-expected decay, we may expect
bursts whose prompt emission is relatively faint compared to the afterglow
emission to show little evidence of energy injection (unless the central engine
gets brighter with time!). The red points in panels e)--f) of
Fig.~\ref{fig:closure}, and the points in panel g) are such bursts: their
afterglows show no steep-decay phase, which (see Section~\ref{sec:discuss}) may
mean that from an early time, the afterglow dominated any prompt emission. As
can be seen, the majority of these are consistent with having no
energy injection, supporting this model.

The discussion above does not tell us that the forward shock model for X-ray
afterglows is the correct model for XRT afterglow emission, however it
demonstrated that, with a little reorganising in light of the constraints placed
by our dataset, it is still consistent with observations. Nonetheless, two
difficulties remain: some mechanism must be found to produce a spectrally
invariant temporal break with a wide range of $\Delta\alpha$; and it must be
possible to inject a significant amount of energy into the external shock for
days to weeks after the explosion.

\section{Conclusions}

We have developed software to automatically produce light curves and hardness
ratios, spectra and high-precision enhanced XRT positions of GRBs. Preliminary
versions of these are available within minutes of a trigger, and the full
versions available within a few hours. Users can interact with and customise
these products as desired. We also provide general purpose versions of these
tools to run for any object observed by XRT, available via a web interface.

These products are available online:

\begin{itemize}
\item{Index: http://www.swift.ac.uk/xrt\_products}
\item{Positions: http://www.swift.ac.uk/xrt\_positions}
\item{Light curves: http://www.swift.ac.uk/xrt\_curves}
\item{Spectra: http://www.swift.ac.uk/xrt\_spectra}
\end{itemize}

Using this software we have performed a homogeneous analysis of all GRBs
observed by the XRT to date, and presented positions and temporal and spectral
indices, in various formats. Analysis of these data show that a variety of
light curve morphologies exist, and the so-called `canonical' curve, while the
most common case, accounts for less than half of the light curves seen by
\swift. Defining a subsample of 162 GRBs with sufficient coverage to detect the
canonical shape, if it existed, we found:

\begin{itemize}
\item{8 (5 \%) have no breaks.}
\item{49 (30 \%) have one break (25 shallow, 24 steepen).}
\item{67 (41 \%) are canonical.}
\item{38 (24 \%) are oddballs.}
\end{itemize}

We have, however, demonstrated that this range of morphologies can be
explained by a single underlying behaviour; the two-component model suggested by
Willingale \etal(2007), which involves a `prompt' component and an `afterglow'
component. To achieve this we require a range of prompt-to-afterglow emission
ratios, and a range of energy injection rates, both of which are easy to accept
given the variations seen from burst to burst. 

If the afterglow emission is due to the external forward shock model then in
many cases this scenario  can only explain the data if energy injection
continues beyond the plateau phase, and lasts for days to weeks after the GRB.
The data also require a mechanism which can cause a light curve break (i.e.\ the
end of the plateau) without terminating energy injection, and without causing a
change in the X-ray spectrum.

\subsection{Usage policy}
\label{sec:usage}

Anybody is welcome to use the products and tools details in this paper for their work.
Although we have verified these tools as far as possible, we still strongly
advise users to `sanity check' their results, particularly with regards to light
curve binning (Section~\ref{sec:ngrblc}).

If these products or tools are used in any publication, we ask that this paper
be cited, and that users include the following statement in the
acknowledgements:

``This work made use of data supplied by the UK Swift Science Data Centre at the
University of Leicester.''

\clearpage

\begin{table}
\begin{center}
\begin{tabular}{ll}
\hline
GRB  &  Times ignored from fit \\
     & (s since trigger) \\
\hline
GRB 050406  &  115--430 \\
GRB 050410  &  45000--100000 \\
GRB 050502B  &  298--1680, 24800--120000 \\
GRB 050607  &  223--428 \\
GRB 050712  &  179--960 \\
GRB 050713A  &  96--152, 157--203 \\
\end{tabular}
\caption{Times identified as deviating from power-law decays, and excluded from the light
curve fits, see Section~\ref{sec:flares} for details.}
\label{tab:flares}
\end{center}
\end{table}

%\addtocounter{table}{-1}
\begin{table}
\begin{center}
\begin{tabular}{ll}
\hline
GRB  &  Times ignored from fit \\
     & (s since trigger) \\
\hline
GRB 070616  &  480--550, 714--995 \\
GRB 070621  &  135--220 \\
GRB 070704  &  257--658 \\
GRB 070721B  &  124--141, 187--858 \\
GRB 070724A  &  90--150 \\
GRB 071021  &  148--300, 1000--9000 \\
GRB 071031  &  109--700 \\
GRB 071112C  &  523--726 \\
GRB 071118  &  189--225, 321--900 \\
GRB 071122  &  398--498 \\
GRB 080210  &  165--379 \\
GRB 080212  &  169--491 \\
GRB 080229A  &  100--156 \\
GRB 080310  &  130--980 \\
GRB 080319D  &  264--640 \\
GRB 080320  &  245--500, 650--1000 \\
GRB 080325  &  158--400 \\
GRB 080506  &  364--1200 \\
GRB 080604  &  18000--100000 \\
GRB 080607  &  116--210, 3000--7000 \\
\end{tabular}
\contcaption{}
\end{center}
\end{table}

\pagebreak

%%%%%%%%%%%%%%%%%%%%%%
%%% START OF POS TABLE
%%%%%%%%%%%%%%%%%%%%%%
\begin{table}
\begin{center}
\begin{tabular}{lccc}
\hline
GRB  &  RA (J2000.0)  & Dec (J2000.0) & Error$^1$ \\
\hline
GRB050124  &  12 51 30.55  &  +13 02 39.6  &  1.6 \\
GRB050128  &  14 38 17.73  &  $-$34 45 55.2  &  1.7 \\
GRB050219A  &  11 05 39.01  &  $-$40 41 03.1  &  2.1 \\
GRB050219B  &  05 25 16.05  &  $-$57 45 29.3  &  1.5 \\
GRB050223  &  18 05 32.38  &  $-$62 28 21.9  &  4.3 \\
GRB050315  &  20 25 54.20  &  $-$42 36 01.5  &  1.5 \\
GRB050318  &  03 18 50.99  &  $-$46 23 45.0  &  1.4 \\
\hline
\end{tabular}
\caption{All available enhanced XRT positions of GRBs observed by \swift. 
The positions are also available online at http://www.swift.ac.uk/xrt\_positions, 
which is updated automatically
with new GRBs.
\newline $^1$arc seconds, 90\%\ confidence radius. }
\label{tab:enhpos}
\end{center}
\end{table}

\begin{table}
\begin{center}
\begin{tabular}{lccc}
\hline
GRB  &  RA (J2000.0)  & Dec (J2000.0) & Error$^1$ \\
\hline
GRB051117B  &  05 40 43.39  &  $-$19 16 26.9  &  1.7 \\
GRB051210  &  22 00 41.33  &  $-$57 36 49.4  &  1.7 \\
GRB051211B  &  23 02 41.53  &  +55 04 50.7  &  1.6 \\
GRB051221A  &  21 54 48.58  &  +16 53 26.0  &  1.4 \\
GRB051221B  &  20 49 35.25  &  +53 02 10.7  &  2.1 \\
GRB051227  &  08 20 58.31  &  +31 55 31.0  &  1.6 \\
GRB060105  &  19 50 00.68  &  +46 20 55.4  &  1.4 \\
\hline
\end{tabular}
\contcaption{$^1$arc seconds, 90\%\ confidence radius. Also
available online at http://www.swift.ac.uk/xrt\_positions, updated automatically
with new GRBs.}
\end{center}
\end{table}

\begin{table}
\begin{center}
\begin{tabular}{lccc}
\hline
GRB  &  RA (J2000.0)  & Dec (J2000.0) & Error$^1$ \\
\hline
GRB060814  &  14 45 21.32  &  +20 35 09.2  &  1.4 \\
GRB060825  &  01 12 29.43  &  +55 47 51.8  &  2.0 \\
GRB060901  &  19 08 37.81  &  $-$06 38 05.7  &  2.9 \\
GRB060904B  &  03 52 50.56  &  $-$00 43 30.1  &  1.4 \\
GRB060906  &  02 43 00.90  &  +30 21 43.0  &  1.4 \\
GRB060908  &  02 07 18.42  &  +00 20 32.4  &  1.4 \\
GRB060912A  &  00 21 08.11  &  +20 58 19.2  &  1.4 \\
\hline
\end{tabular}
\contcaption{$^1$arc seconds, 90\%\ confidence radius. Also
available online at http://www.swift.ac.uk/xrt\_positions, updated automatically
with new GRBs.}
\end{center}
\end{table}

\begin{table}
\begin{center}
\begin{tabular}{lccc}
\hline
GRB  &  RA (J2000.0)  & Dec (J2000.0) & Error$^1$ \\
\hline
GRB070611  &  00 07 58.15  &  $-$29 45 20.4  &  1.8 \\
GRB070612B  &  17 26 54.46  &  $-$08 45 04.7  &  1.8 \\
GRB070616  &  02 08 36.40  &  +56 56 45.0  &  1.4 \\
GRB070621  &  21 35 10.17  &  $-$24 49 02.6  &  1.4 \\
GRB070628  &  07 41 06.06  &  $-$20 16 45.3  &  1.4 \\
GRB070704  &  23 38 47.31  &  +66 15 12.7  &  1.7 \\
GRB070714A  &  02 51 43.36  &  +30 14 35.1  &  1.8 \\
\hline
\end{tabular}
\contcaption{$^1$arc seconds, 90\%\ confidence radius. Also
available online at http://www.swift.ac.uk/xrt\_positions, updated automatically
with new GRBs.}
\end{center}
\end{table}

\begin{table}
\begin{center}
\begin{tabular}{lccc}
\hline
GRB  &  RA (J2000.0)  & Dec (J2000.0) & Error$^1$ \\
\hline
GRB080605  &  17 28 30.04  &  +04 00 56.9  &  1.5 \\
GRB080607  &  12 59 47.14  &  +15 55 09.6  &  1.4 \\
GRB080613A  &  14 13 04.98  &  +05 10 21.8  &  2.0 \\
GRB080613B  &  11 35 11.51  &  $-$07 06 17.5  &  2.5 \\
GRB080623  &  15 50 38.65  &  $-$62 02 57.3  &  1.5 \\
GRB080625  &  19 53 37.09  &  +56 16 41.7  &  1.7 \\
GRB080701  &  03 03 21.31  &  +75 28 29.3  &  1.7 \\
GRB080702A  &  20 52 12.21  &  +72 18 46.7  &  1.8 \\
GRB080702B  &  23 42 32.07  &  $-$05 31 51.7  &  2.2 \\
GRB080703  &  06 47 12.43  &  $-$63 13 08.3  &  1.4 \\
GRB080707  &  02 10 28.48  &  +33 06 35.1  &  1.5 \\
GRB080710  &  00 33 05.68  &  +19 30 05.8  &  1.4 \\
GRB080714  &  12 32 25.29  &  $-$60 16 40.2  &  1.5 \\
GRB080721  &  14 57 55.76  &  $-$11 43 24.9  &  1.4 \\
GRB080723A  &  18 17 57.70  &  $-$06 53 58.3  &  1.4 \\
GRB080723B  &  11 47 19.77  &  $-$60 14 27.9  &  1.5 \\
\hline
\end{tabular}
\contcaption{$^1$arc seconds, 90\%\ confidence radius. Also
available online at http://www.swift.ac.uk/xrt\_positions, updated automatically
with new GRBs.}
\end{center}
\end{table}

\clearpage
%%%%%%%%%%%%%%%%%%%%%%
%%% END OF POS TABLE
%%%%%%%%%%%%%%%%%%%%%%

%%%%%%%%%%%%%%%%%%%%
%%% START OF ALPHA TABLE
%%%%%%%%%%%%%%%%%%%%
\begin{table*}
\begin{center}
\begin{tabular}{lccccccccc}
\hline
GRB  & Target ID & $T_{90}$  &  F$_{11}$  & $\alpha_1$ & $\alpha_2$ & $\alpha_3$ & $\alpha_4$ & $\alpha_5$ & $\alpha_6$ \\
\hline
GRB 041223    &    00100585   &    109.1    &    &   $1.90^{+0.40}_{-0.39}$ \\ 
GRB 050124    &    00103647   &    4.0    &  1.12e-12  &   $1.50^{+0.15}_{-0.15}$ \\ 
GRB 050126    &    00103780   &    24.8    &  1.22e-13  &   $2.57^{+0.99}_{-0.46}$    &    $0.94^{+0.24}_{-0.31}$ \\ 
GRB 050128    &    00103906   &    19.2    &  2.25e-12  &   $0.955^{+0.038}_{-0.108}$    &    $1.361^{+0.075}_{-0.059}$ \\ 
GRB 050215B    &    00106107   &    8.1    &  3.10e-13  &   $0.92^{+0.15}_{-0.13}$ \\ 
GRB 050219A    &    00106415   &    23.7    &  8.73e-13  &   $2.91^{+0.25}_{-0.25}$    &    $0.754^{+0.089}_{-0.083}$ \\ 
GRB 050219B    &    00106442   &    30.7    &  3.53e-12  &   $1.254^{+0.038}_{-0.037}$ \\
\hline
\end{tabular}
\caption{Power-law decay indices from light curve fits. A positive value of
$\alpha$ indicates a decay. The break times between indices are given in
Table~\ref{tab:breaks}. For bursts where the light curve straddles $T0+11$ hrs, 
the model flux at 11 hours is given (in erg \cms\ s$^{-1})$). The \swift\ target ID and BAT $T_{90}$ (taken from the Swift Data
Table) are given for reference.}
\label{tab:alphas}
\end{center}
\end{table*}
\clearpage

\begin{table*}
\begin{center}
\begin{tabular}{lccccccccc}
\hline
GRB  & Target ID & $T_{90}$  &  F$_{11}$  & $\alpha_1$ & $\alpha_2$ & $\alpha_3$ & $\alpha_4$ & $\alpha_5$ & $\alpha_6$ \\
\hline
GRB 051006    &    00158593   &    34.8    &  2.91e-14  &   $1.608^{+0.073}_{-0.063}$ \\ 
GRB 051008    &    00158855   &    $>$32    &  9.91e-13  &   $0.966^{+0.081}_{-0.085}$    &    $2.064^{+0.110}_{-0.097}$ \\ 
GRB 051016A    &    00159913   &    23.0    &  9.66e-14  &   $3.86^{+0.67}_{-0.43}$    &    $0.894^{+0.049}_{-0.047}$ \\ 
GRB 051016B    &    00159994   &    4.0    &  7.22e-13  &   $7.10^{+0.50}_{-1.23}$    &    $0.143^{+0.076}_{-0.086}$    &    $1.009^{+0.049}_{-0.048}$ \\ 
GRB 051021A    &    00020018   &    N/A    &  8.62e-13  &   $1.058^{+0.072}_{-0.063}$ \\ 
GRB 051021B    &    00160672   &    46.5    &  8.94e-14  &   $1.933^{+0.104}_{-0.098}$    &    $0.73^{+0.45}_{-0.56}$ \\ 
GRB 051022    &    00020019   &    N/A    &  8.65e-12  &   $1.547^{+0.075}_{-0.072}$    &    $2.27^{+0.45}_{-0.34}$ \\
\hline
\end{tabular}
\contcaption{}
\end{center}
\end{table*}
\clearpage

\begin{table*}
\begin{center}
\begin{tabular}{lccccccccc}
\hline
GRB  & Target ID & $T_{90}$  &  F$_{11}$  & $\alpha_1$ & $\alpha_2$ & $\alpha_3$ & $\alpha_4$ & $\alpha_5$ & $\alpha_6$ \\
\hline
GRB 060515    &    00210084   &    52.0    &  2.70e-14  &   $1.79^{+0.52}_{-0.38}$ \\ 
GRB 060522    &    00211117   &    71.1    &  1.81e-13  &   $3.80^{+0.63}_{-0.49}$    &    $1.78^{+0.21}_{-0.21}$    &    $0.904^{+0.069}_{-0.054}$ \\ 
GRB 060526    &    00211957   &    298.2    &  8.15e-13  &   $1.222^{+0.073}_{-0.081}$    &    $0.36^{+0.15}_{-0.39}$    &    $0.947^{+0.161}_{-0.085}$    &    $2.91^{+2.19}_{-0.78}$ \\ 
GRB 060602A    &    00213180   &    75.0    &    &   $1.14^{+0.99}_{-0.32}$ \\ 
GRB 060602B    &    00213190   &    N/A    &  2.12e-13  &   $1.115^{+0.102}_{-0.086}$ \\ 
GRB 060604    &    00213486   &    95.0    &  9.56e-13  &   $2.99^{+0.41}_{-0.45}$    &    $0.53^{+0.12}_{-0.47}$    &    $1.238^{+0.080}_{-0.073}$ \\ 
GRB 060605    &    00213630   &    79.1    &  3.81e-13  &   $1.125^{+0.045}_{-0.199}$    &    $0.262^{+0.045}_{-0.279}$    &    $1.96^{+0.11}_{-0.10}$ \\
\hline
\end{tabular}
\contcaption{}
\end{center}
\end{table*}
\clearpage

\begin{table*}
\begin{center}
\begin{tabular}{lccccccccc}
\hline
GRB  & Target ID & $T_{90}$  &  F$_{11}$  & $\alpha_1$ & $\alpha_2$ & $\alpha_3$ & $\alpha_4$ & $\alpha_5$ & $\alpha_6$ \\
\hline
GRB 070103    &    00254532   &    18.6    &  1.62e-13  &   $-0.64^{+0.22}_{-0.37}$    &    $0.86^{+0.18}_{-0.67}$    &    $1.43^{+0.12}_{-0.11}$ \\ 
GRB 070107    &    00255029   &    347.3    &  2.29e-12  &   $0.864^{+0.020}_{-0.021}$    &    $1.308^{+0.046}_{-0.040}$ \\ 
GRB 070110    &    00255445   &    88.4    &  6.83e-13  &   $2.558^{+0.051}_{-0.128}$    &    $0.612^{+0.586}_{-0.095}$    &    $0.132^{+0.037}_{-0.041}$    &    $9.65^{+0.35}_{-0.39}$    &    $0.892^{+0.045}_{-0.040}$ \\ 
GRB 070125    &    00020047   &    N/A    &    &   $0.65^{+0.36}_{-0.27}$    &    $2.01^{+0.18}_{-0.15}$ \\ 
GRB 070129    &    00258408   &    460.6    &  1.14e-12  &   $2.532^{+0.079}_{-0.070}$    &    $-0.093^{+0.230}_{-0.091}$    &    $1.034^{+0.051}_{-0.049}$    &    $2.41^{+1.38}_{-0.97}$ \\ 
GRB 070208    &    00259714   &    47.7    &  3.42e-13  &   $-0.98^{+0.20}_{-0.23}$    &    $0.99^{+0.13}_{-0.38}$    &    $1.58^{+0.59}_{-0.28}$ \\ 
GRB 070219    &    00261132   &    16.6    &  1.02e-13  &   $1.67^{+0.94}_{-0.55}$    &    $-0.27^{+0.75}_{-0.80}$    &    $0.99^{+0.20}_{-0.17}$ \\
\hline
\end{tabular}
\contcaption{}
\end{center}
\end{table*}
\clearpage

\begin{table*}
\begin{center}
\begin{tabular}{lccccccccc}
\hline
GRB  & Target ID & $T_{90}$  &  F$_{11}$  & $\alpha_1$ & $\alpha_2$ & $\alpha_3$ & $\alpha_4$ & $\alpha_5$ & $\alpha_6$ \\
\hline
GRB 071021    &    00294974   &    225    &  3.10e-13  &   $4.40^{+0.43}_{-0.26}$    &    $0.813^{+0.042}_{-0.044}$ \\ 
GRB 071025    &    00295301   &    109    &  5.72e-13  &   $3.59^{+0.18}_{-0.26}$    &    $2.25^{+0.14}_{-0.20}$    &    $1.322^{+0.054}_{-0.040}$    &    $1.668^{+0.064}_{-0.063}$ \\ 
GRB 071028A    &    00295527   &    27.0    &  1.83e-13  &   $3.18^{+0.19}_{-0.17}$    &    $-0.06^{+0.21}_{-0.59}$    &    $1.95^{+1.20}_{-0.85}$ \\ 
GRB 071031    &    00295670   &    18    &  2.51e-13  &   $4.84^{+0.48}_{-0.31}$    &    $0.95^{+0.14}_{-0.39}$    &    $1.30^{+8.70}_{-0.35}$ \\ 
GRB 071101    &    00295779   &    9.0    &  6.64e-14  &   $0.73^{+0.13}_{-0.15}$ \\ 
GRB 071104    &    00020058   &    N/A    &  8.88e-13  &   $1.34^{+0.26}_{-0.26}$ \\ 
GRB 071112C    &    00296504   &    15    &  3.21e-13  &   $1.424^{+0.096}_{-0.117}$    &    $0.81^{+0.13}_{-0.18}$    &    $1.535^{+0.048}_{-0.045}$ \\
\hline
\end{tabular}
\contcaption{}
\end{center}
\end{table*}
\clearpage

\begin{table*}
\begin{center}
\begin{tabular}{lccccccccc}
\hline
GRB  & Target ID & $T_{90}$  &  F$_{11}$  & $\alpha_1$ & $\alpha_2$ & $\alpha_3$ & $\alpha_4$ & $\alpha_5$ & $\alpha_6$ \\
\hline
GRB 080625    &    00020076   &    N/A    &  1.88e-12  &   $1.409^{+0.085}_{-0.169}$    &    $2.39^{+1.29}_{-0.61}$ \\ 
GRB 080701    &    00315615   &    18    &  3.52e-13  &   $1.07^{+0.22}_{-0.15}$ \\ 
GRB 080702A    &    00315710   &    0.5    &  3.31e-14  &   $1.054^{+0.116}_{-0.095}$ \\ 
GRB 080703    &    00315819   &    3.4    &  8.08e-13  &   $0.559^{+0.036}_{-0.037}$    &    $2.30^{+0.29}_{-0.24}$ \\ 
GRB 080707    &    00316204   &    27.1    &  4.68e-13  &   $4.39^{+1.10}_{-0.57}$    &    $0.21^{+0.11}_{-0.12}$    &    $1.048^{+0.098}_{-0.092}$ \\ 
GRB 080710    &    00316534   &    12    &  8.31e-13  &   $1.050^{+0.081}_{-0.063}$    &    $1.80^{+0.23}_{-0.17}$ \\ 
GRB 080714    &    00316910   &    33    &  6.44e-13  &   $0.10^{+0.64}_{-0.92}$    &    $1.87^{+2.62}_{-0.49}$    &    $1.025^{+0.049}_{-0.047}$ \\ 
GRB 080721    &    00317508   &    16.2    &  7.26e-12  &   $0.710^{+0.046}_{-0.021}$    &    $-1.10^{+1.34}_{-0.40}$    &    $0.938^{+0.020}_{-0.020}$    &    $1.639^{+0.022}_{-0.022}$ \\ 
GRB 080723A    &    00317662   &    17.3    &  9.78e-13  &   $1.82^{+0.36}_{-0.31}$    &    $0.02^{+0.12}_{-0.25}$    &    $1.060^{+0.059}_{-0.054}$ \\ 
GRB 080723B    &    00020079   &    N/A    &  6.66e-12  &   $1.40^{+0.19}_{-0.20}$ \\ 
\hline
\end{tabular}
\contcaption{}
\end{center}
\end{table*}
\clearpage

%%%%%%%%%%%%%%%%%%%%
%%% END OF ALPHA TABLE
%%%%%%%%%%%%%%%%%%%%

%%%%%%%%%%%%%%%%%%%%
%%% START OF BREAK TABLE
%%%%%%%%%%%%%%%%%%%%

\begin{table*}
\begin{center}
\begin{tabular}{lcccccc}
\hline
GRB  & Obs times$^*$ & $t{_{\rm break},1}$ & $t{_{\rm break},2}$ & $t{_{\rm break},3}$ & $t{_{\rm break},4}$ & $t{_{\rm break},5}$\\
\hline
GRB 041223 & 16.7--28.6 ks  \\
GRB 050124 & 11.1--4967.5 ks  \\
GRB 050126 & 128 s--93.2 ks   &  $495^{+540}_{-260}$ \\
GRB 050128 & 105 s--99.6 ks   &  $6492^{+2247}_{-3355}$ \\
GRB 050215B & 5.8--3011.3 ks  \\
GRB 050219A & 112 s--3154.9 ks   &  $339^{+53}_{-37}$ \\
GRB 050219B & 3.2--3205.2 ks  \\
\hline
\end{tabular}
\caption{Times of the breaks in the light curve fits. The power-law indices are
given in Table~\ref{tab:alphas}.
\newline $^*$ Time range covered by the cleaned event lists. Zero is the
BAT trigger time.}
\label{tab:breaks}
\end{center}
\end{table*}
\clearpage

\begin{table*}
\begin{center}
\begin{tabular}{lcccccc}
\hline
GRB  & Obs times$^*$ & $t{_{\rm break},1}$ & $t{_{\rm break},2}$ & $t{_{\rm break},3}$ & $t{_{\rm break},4}$ & $t{_{\rm break},5}$\\
\hline
GRB 051006 & 113 s--358.1 ks  \\
GRB 051008 & 3.0--542.4 ks   &  $(1.74^{+0.23}_{+0.20}) \tim{4}$ \\
GRB 051016A & 120 s--931.0 ks   &  $291^{+39}_{-41}$ \\
GRB 051016B & 83 s--1452.0 ks   &  $152^{+12}_{-9}$  &  $7103^{+1488}_{-1523}$ \\
GRB 051021A & 11.2--1430.5 ks  \\
GRB 051021B & 85 s--516.4 ks   &  $2584^{+2972}_{-1497}$ \\
GRB 051022 & 12.5--1263.3 ks   &  $(2.50^{+0.61}_{+0.55}) \tim{5}$ \\
\hline
\end{tabular}
\contcaption{$^*$ Time range covered by the cleaned event lists. Zero is the
BAT trigger time.}
\end{center}
\end{table*}
\clearpage

\begin{table*}
\begin{center}
\begin{tabular}{lcccccc}
\hline
GRB  & Obs times$^*$ & $t{_{\rm break},1}$ & $t{_{\rm break},2}$ & $t{_{\rm break},3}$ & $t{_{\rm break},4}$ & $t{_{\rm break},5}$\\
\hline
GRB 060515 & 3.4--40.6 ks  \\
GRB 060522 & 147 s--509.1 ks   &  $226^{+37}_{-19}$  &  $927^{+236}_{-130}$ \\
GRB 060526 & 81 s--545.5 ks   &  $1433^{+1060}_{-295}$  &  $6592^{+9496}_{-4265}$  &  $(1.23^{+0.52}_{+0.33}) \tim{5}$ \\
GRB 060602A & 152.6--870.0 ks  \\
GRB 060602B & 88 s--653.7 ks  \\
GRB 060604 & 116 s--1920.6 ks   &  $635^{+172}_{-134}$  &  $(2.50^{+0.87}_{+1.26}) \tim{4}$ \\
GRB 060605 & 98 s--190.9 ks   &  $535^{+434}_{-161}$  &  $7465^{+597}_{-632}$ \\
\hline
\end{tabular}
\contcaption{$^*$ Time range covered by the cleaned event lists. Zero is the
BAT trigger time.}
\end{center}
\end{table*}
\clearpage

\begin{table*}
\begin{center}
\begin{tabular}{lcccccc}
\hline
GRB  & Obs times$^*$ & $t{_{\rm break},1}$ & $t{_{\rm break},2}$ & $t{_{\rm break},3}$ & $t{_{\rm break},4}$ & $t{_{\rm break},5}$\\
\hline
GRB 070103 & 72 s--179.9 ks   &  $616^{+55}_{-180}$  &  $4628^{+1985}_{-2918}$ \\
GRB 070107 & 182 s--817.1 ks   &  $(1.76^{+0.39}_{+0.28}) \tim{4}$ \\
GRB 070110 & 99 s--2306.2 ks   &  $317^{+45}_{-42}$  &  $4043^{+440}_{-2940}$  &  $(2.068^{+0.021}_{+0.021}) \tim{4}$  &  $(2.716^{+0.082}_{+0.041}) \tim{4}$ \\
GRB 070125 & 46.7--1614.9 ks   &  $(10^{+4}_{+2}) \tim{4}$ \\
GRB 070129 & 139 s--1637.1 ks   &  $1786^{+284}_{-179}$  &  $(1.21^{+0.17}_{+0.22}) \tim{4}$  &  $(8^{+5}_{+4}) \tim{5}$ \\
GRB 070208 & 119 s--480.4 ks   &  $923^{+351}_{-277}$  &  $(1.43^{+1.32}_{+0.59}) \tim{4}$ \\
GRB 070219 & 86 s--255.0 ks   &  $627^{+1167}_{-232}$  &  $3315^{+53930}_{-1827}$ \\
\hline
\end{tabular}
\contcaption{$^*$ Time range covered by the cleaned event lists. Zero is the
BAT trigger time.}
\end{center}
\end{table*}
\clearpage

\begin{table*}
\begin{center}
\begin{tabular}{lcccccc}
\hline
GRB  & Obs times$^*$ & $t{_{\rm break},1}$ & $t{_{\rm break},2}$ & $t{_{\rm break},3}$ & $t{_{\rm break},4}$ & $t{_{\rm break},5}$\\
\hline
GRB 071021 & 136 s--1000.7 ks   &  $447^{+51}_{-60}$ \\
GRB 071025 & 149 s--497.6 ks   &  $183^{+15}_{-10}$  &  $318^{+305}_{-18}$  &  $3557^{+1494}_{-3061}$ \\
GRB 071028A & 124 s--87.5 ks   &  $1228^{+529}_{-277}$  &  $(2.52^{+0.80}_{+1.03}) \tim{4}$ \\
GRB 071031 & 109 s--596.7 ks   &  $1608^{+230}_{-450}$  &  $(6^{+53}_{+6}) \tim{4}$ \\
GRB 071101 & 72 s--92.7 ks  \\
GRB 071104 & 20.6--67.1 ks  \\
GRB 071112C & 89 s--624.4 ks   &  $243^{+71}_{-35}$  &  $881^{+176}_{-159}$ \\
\hline
\end{tabular}
\contcaption{$^*$ Time range covered by the cleaned event lists. Zero is the
BAT trigger time.}
\end{center}
\end{table*}
\clearpage

\begin{table*}
\begin{center}
\begin{tabular}{lcccccc}
\hline
GRB  & Obs times$^*$ & $t{_{\rm break},1}$ & $t{_{\rm break},2}$ & $t{_{\rm break},3}$ & $t{_{\rm break},4}$ & $t{_{\rm break},5}$\\
\hline
GRB 080625 & 17.9--818.1 ks   &  $(1.67^{+0.88}_{+1.09}) \tim{5}$ \\
GRB 080701 & 2.9--131.2 ks  \\
GRB 080702A & 74 s--121.6 ks  \\
GRB 080703 & 103 s--180.9 ks   &  $(2.14^{+0.23}_{+0.23}) \tim{4}$ \\
GRB 080707 & 74 s--416.6 ks   &  $184^{+18}_{-26}$  &  $7703^{+2997}_{-6696}$ \\
GRB 080710 & 3.1--646.9 ks   &  $(2.21^{+0.55}_{+0.52}) \tim{4}$ \\
GRB 080714 & 86 s--450.2 ks   &  $126^{+71}_{-27}$  &  $287^{+158}_{-72}$ \\
GRB 080721 & 113 s--1405.7 ks   &  $290^{+9}_{-27}$  &  $306^{+12}_{-9}$  &  $3990^{+346}_{-323}$ \\
GRB 080723A & 96 s--573.3 ks   &  $380^{+92}_{-71}$  &  $2880^{+1151}_{-613}$ \\
GRB 080723B & 13.4--40.6 ks  \\
\hline
\end{tabular}
\contcaption{$^*$ Time range covered by the cleaned event lists. Zero is the
BAT trigger time.}
\end{center}
\end{table*}
\clearpage

%%%%%%%%%%%%%%%%%%%%
%%% END OF BREAK TABLE
%%%%%%%%%%%%%%%%%%%%

%%%%%%%%%%%%%%%%%%%%
%%% START OF SPEC TABLE
%%%%%%%%%%%%%%%%%%%%
\clearpage

\begin{table*}
\begin{center}
\begin{tabular}{lcccc|cc}
\hline
%     & \colspan{2}{WT mode} & \colspan{2}{PC mode} \\
     &                  &            & \multicolumn{2}{c}{WT Mode} & \multicolumn{2}{c}{PC Mode} \\
GRB  &  Galactic $N_{\rm H}$  &  redshift  &  intrinsic $N_{\rm H}$  &  Spectral index ($\beta)$  &  intrinsic $N_{\rm H}$  &  Spectral index ($\beta)$ \\
\hline
GRB 041218  &  27.4  &    &  $21.6^{+133.6}_{-21.6}$  &  $2.3^{+5.4}_{-2.0}$  &  $<50.27$  &  $0.55^{+1.23}_{-0.69}$  \\
GRB 041223  &  10.1  &    &    &  &  $3.1^{+4.5}_{-3.1}$  &  $1.13^{+0.19}_{-0.16}$  \\
GRB 050124  &  2.66  &    &  $0.96^{+8.00}_{-0.96}$  &  $0.54^{+0.35}_{-0.24}$  &  $2.1^{+2.6}_{-2.0}$  &  $0.94^{+0.21}_{-0.20}$  \\
GRB 050126  &  4.62  &  1.29  &    &  &  $<11.50$  &  $1.14^{+0.22}_{-0.15}$  \\
GRB 050128  &  5.19  &    &  $<125.82$  &  $0.41^{+1.80}_{-0.45}$  &  $4.9^{+2.0}_{-1.9}$  &  $0.968^{+0.087}_{-0.044}$  \\
GRB 050215B  &  1.85  &    &    &  &  $<8.82$  &  $0.65^{+0.59}_{-0.28}$  \\
\hline
\end{tabular}
\caption{Best-fit results for time-averaged spectra. Column densities are in units of $10^{20}$ \cms. WT and PC mode were fitted independently.}
\label{tab:avespec}
\end{center}
\end{table*}
\clearpage

%\addtocounter{table}{-1}
\begin{table*}
\begin{center}
\begin{tabular}{lcccc|cc}
\hline
%     & \colspan{2}{WT mode} & \colspan{2}{PC mode} \\
     &                  &            & \multicolumn{2}{c}{WT Mode} & \multicolumn{2}{c}{PC Mode} \\
GRB  &  Galactic $N_{\rm H}$  &  redshift  &  intrinsic $N_{\rm H}$  &  Spectral index ($\beta)$  &  intrinsic $N_{\rm H}$  &  Spectral index ($\beta)$ \\
\hline
GRB 050908  &  2.38  &  3.35  &    &  &  $0.65^{+44.49}_{-0.65}$  &  $1.91^{+0.24}_{-0.13}$  \\
GRB 050915A  &  1.97  &    &  $11.5^{+5.7}_{-5.2}$  &  $1.18^{+0.27}_{-0.25}$  &  $9.2^{+3.5}_{-4.5}$  &  $1.29^{+0.16}_{-0.22}$  \\
GRB 050915B  &  22.2  &    &  $9.7^{+4.9}_{-4.5}$  &  $1.68^{+0.20}_{-0.19}$  &  $18.4^{+7.8}_{-10.3}$  &  $1.36^{+0.19}_{-0.27}$  \\
GRB 050916  &  81.0  &    &    &  &  $13.5^{+19.4}_{-12.2}$  &  $0.64^{+0.16}_{-0.23}$  \\
GRB 050922B  &  3.14  &    &  $13.3^{+1.6}_{-1.5}$  &  $1.700^{+0.086}_{-0.083}$  &  $13.4^{+2.7}_{-2.5}$  &  $1.58^{+0.13}_{-0.12}$  \\
GRB 050922C  &  5.37  &  2.198  &  $17.3^{+14.4}_{-13.7}$  &  $1.028^{+0.073}_{-0.072}$  &  $43.7^{+18.3}_{-10.6}$  &  $1.287^{+0.084}_{-0.077}$  \\
\hline
\end{tabular}
\end{center}
\end{table*}
\clearpage

%\addtocounter{table}{-1}
\begin{table*}
\begin{center}
\begin{tabular}{lcccc|cc}
\hline
%     & \colspan{2}{WT mode} & \colspan{2}{PC mode} \\
     &                  &            & \multicolumn{2}{c}{WT Mode} & \multicolumn{2}{c}{PC Mode} \\
GRB  &  Galactic $N_{\rm H}$  &  redshift  &  intrinsic $N_{\rm H}$  &  Spectral index ($\beta)$  &  intrinsic $N_{\rm H}$  &  Spectral index ($\beta)$ \\
\hline
GRB 060427  &  3.85  &    &  $10.3^{+3.0}_{-2.8}$  &  $1.78^{+0.18}_{-0.17}$  &  $12.0^{+4.7}_{-4.9}$  &  $1.42^{+0.21}_{-0.22}$  \\
GRB 060428A  &  76.7  &    &  $<16.19$  &  $2.30^{+0.53}_{-0.26}$  &  $18.6^{+12.6}_{-11.7}$  &  $1.03^{+0.15}_{-0.15}$  \\
GRB 060428B  &  1.33  &    &  $6.5^{+1.5}_{-1.4}$  &  $1.86^{+0.11}_{-0.10}$  &  $<1.28$  &  $1.040^{+0.115}_{-0.090}$  \\
GRB 060501  &  22.8  &    &    &  &  $30.0^{+99.4}_{-30.0}$  &  $1.62^{+2.21}_{-0.57}$  \\
GRB 060502A  &  3.45  &  1.51  &  $53.8^{+16.8}_{-15.3}$  &  $2.48^{+0.27}_{-0.25}$  &  $24.1^{+13.5}_{-12.8}$  &  $1.034^{+0.115}_{-0.058}$  \\
GRB 060502B  &  4.24  &    &    &  &  $7.6^{+22.3}_{-7.6}$  &  $1.10^{+2.77}_{-0.81}$  \\
\hline
\end{tabular}
\end{center}
\end{table*}

\clearpage

%\addtocounter{table}{-1}
\begin{table*}
\begin{center}
\begin{tabular}{lcccc|cc}
\hline
%     & \colspan{2}{WT mode} & \colspan{2}{PC mode} \\
     &                  &            & \multicolumn{2}{c}{WT Mode} & \multicolumn{2}{c}{PC Mode} \\
GRB  &  Galactic $N_{\rm H}$  &  redshift  &  intrinsic $N_{\rm H}$  &  Spectral index ($\beta)$  &  intrinsic $N_{\rm H}$  &  Spectral index ($\beta)$ \\
\hline
GRB 061102  &  4.57  &    &  $8.5^{+5.4}_{-4.7}$  &  $2.20^{+0.40}_{-0.35}$  &  $3.0^{+34.6}_{-3.0}$  &  $0.36^{+0.92}_{-0.70}$  \\
GRB 061110A  &  4.26  &    &  $10.3^{+1.2}_{-1.1}$  &  $1.989^{+0.074}_{-0.068}$  &  $<0.70$  &  $0.86^{+0.15}_{-0.10}$  \\
GRB 061110B  &  3.37  &  3.44  &    &  &  $342.2^{+250.0}_{-257.8}$  &  $1.24^{+0.31}_{-0.27}$  \\
GRB 061121  &  3.99  &  1.314  &  $42.9^{+6.1}_{-5.8}$  &  $0.617^{+0.042}_{-0.041}$  &  $53.5^{+8.5}_{-8.0}$  &  $0.914^{+0.061}_{-0.059}$  \\
GRB 061122  &  12.5  &    &    &  &  $5.1^{+4.7}_{-4.7}$  &  $1.04^{+0.17}_{-0.18}$  \\
GRB 061126  &  10.2  &    &  $9.5^{+4.6}_{-4.2}$  &  $0.86^{+0.16}_{-0.15}$  &  $14.1^{+2.1}_{-2.0}$  &  $0.936^{+0.065}_{-0.064}$  \\
\hline
\end{tabular}
\end{center}
\end{table*}
\clearpage

%\addtocounter{table}{-1}
\begin{table*}
\begin{center}
\begin{tabular}{lcccc|cc}
\hline
%     & \colspan{2}{WT mode} & \colspan{2}{PC mode} \\
     &                  &            & \multicolumn{2}{c}{WT Mode} & \multicolumn{2}{c}{PC Mode} \\
GRB  &  Galactic $N_{\rm H}$  &  redshift  &  intrinsic $N_{\rm H}$  &  Spectral index ($\beta)$  &  intrinsic $N_{\rm H}$  &  Spectral index ($\beta)$ \\
\hline
GRB 070721B  &  2.33  &    &  $1.6^{+2.1}_{-1.6}$  &  $0.421^{+0.091}_{-0.085}$  &  $0.76^{+2.32}_{-0.76}$  &  $0.501^{+0.098}_{-0.050}$  \\
GRB 070724A  &  1.17  &    &  $21.0^{+6.5}_{-6.0}$  &  $0.60^{+0.18}_{-0.18}$  &  $11.6^{+10.4}_{-11.6}$  &  $0.69^{+0.28}_{-0.29}$  \\
GRB 070724B  &  30.7  &    &    &  &  $27.5^{+15.8}_{-17.4}$  &  $1.17^{+0.49}_{-0.40}$  \\
GRB 070729  &  2.58  &    &    &  &  $0.024^{+9.135}_{-0.024}$  &  $0.62^{+0.86}_{-0.43}$  \\
GRB 070802  &  2.90  &  2.45  &    &  &  $122.7^{+80.0}_{-100.7}$  &  $1.08^{+0.17}_{-0.18}$  \\
GRB 070805  &  19.2  &    &    &  &  $0.057^{+39.184}_{-0.057}$  &  $1.2^{+4.0}_{-1.3}$  \\
\hline
\end{tabular}
\end{center}
\end{table*}
\clearpage

%\addtocounter{table}{-1}
\begin{table*}
\begin{center}
\begin{tabular}{lcccc|cc}
\hline
%     & \colspan{2}{WT mode} & \colspan{2}{PC mode} \\
     &                  &            & \multicolumn{2}{c}{WT Mode} & \multicolumn{2}{c}{PC Mode} \\
GRB  &  Galactic $N_{\rm H}$  &  redshift  &  intrinsic $N_{\rm H}$  &  Spectral index ($\beta)$  &  intrinsic $N_{\rm H}$  &  Spectral index ($\beta)$ \\
\hline
GRB 080405  &  3.31  &    &    &  &  $<12.47$  &  $0.89^{+0.52}_{-0.36}$  \\
GRB 080409  &  23.4  &    &    &  &  $37.3^{+14.2}_{-19.1}$  &  $1.28^{+0.30}_{-0.22}$  \\
GRB 080411  &  5.77  &  1.03  &    &  &  $44.4^{+4.8}_{-4.5}$  &  $1.024^{+0.044}_{-0.044}$  \\
GRB 080413A  &  8.71  &  2.43  &  $187.2^{+42.1}_{-37.9}$  &  $2.17^{+0.17}_{-0.16}$  &  $53.4^{+75.8}_{-53.4}$  &  $1.13^{+0.21}_{-0.24}$  \\
GRB 080413B  &  3.06  &  1.1  &  $27.5^{+9.2}_{-8.5}$  &  $1.032^{+0.103}_{-0.100}$  &  $26.0^{+5.4}_{-5.4}$  &  $0.989^{+0.065}_{-0.040}$  \\
GRB 080426  &  36.5  &    &    &  &  $16.5^{+5.1}_{-11.5}$  &  $1.10^{+0.12}_{-0.25}$  \\
\hline
\end{tabular}
\end{center}
\end{table*}

\clearpage

%%%%%%%%%%%%%%%%%%%%
%%% END OF SPEC TABLE
%%%%%%%%%%%%%%%%%%%%

%%%%%%%%%%%%%%%%%%%%
%%% RES BETA TABLE
%%%%%%%%%%%%%%%%%%%%
\clearpage

\begin{table*}
\begin{center}
\begin{tabular}{lccccccc}
\hline
GRB  & Mode & $\beta_1$ & $\beta_2$ & $\beta_3$ & $\beta_4$ & $\beta_5$ & $\beta_6$  \\
\hline
GRB 050126  &  PC  &  1.29$^{+0.20}_{-0.20}$  &  1.04$^{+0.52}_{-0.36}$   \\
GRB 050128  &  WT    &  &  0.35$^{+1.74}_{-0.63}$   \\
&  PC  &  0.98$^{+0.13}_{-0.12}$  &  0.966$^{+0.088}_{-0.092}$   \\
GRB 050219A  &  WT  &  1.08$^{+0.18}_{-0.17}$  &  1.09$^{+0.29}_{-0.13}$   \\
&  PC    &  &  1.16$^{+0.48}_{-0.53}$   \\
GRB 050315  &  WT    &  &  1.09$^{+0.34}_{-0.29}$   \\
&  PC  &  1.07$^{+0.11}_{-0.10}$  &  0.979$^{+0.085}_{-0.059}$  &  1.088$^{+0.060}_{-0.104}$   \\
\hline
\end{tabular}
\caption{Spectral indices for the time-resolved spectra. $\beta_1$ corresponds
to the time during which the decay followed $\alpha_1$ in
Table~6. Column densities for these spectra are given in
Table~10.}
\label{tab:resbeta}
\end{center}
\end{table*}
\clearpage

%\addtocounter{table}{-1}
\begin{table*}
\begin{center}
\begin{tabular}{lccccccc}
\hline
GRB  & Mode & $\beta_1$ & $\beta_2$ & $\beta_3$ & $\beta_4$ & $\beta_5$ & $\beta_6$  \\
\hline
GRB 050819  &  WT  &  1.29$^{+0.27}_{-0.15}$   \\
&  PC  &  1.33$^{+0.32}_{-0.22}$  &  1.40$^{+0.50}_{-0.47}$   \\
GRB 050820A  &  WT  &  1.04$^{+0.20}_{-0.18}$  &  -0.197$^{+0.050}_{-0.038}$   \\
&  PC    &  &  0.853$^{+0.080}_{-0.077}$  &  1.084$^{+0.035}_{-0.034}$   \\
GRB 050822  &  WT  &  1.583$^{+0.074}_{-0.070}$  &  3.42$^{+2.18}_{-0.91}$  &  0.40$^{+1.25}_{-0.54}$   \\
&  PC  &  1.86$^{+0.70}_{-0.44}$  &  1.00$^{+0.17}_{-0.15}$  &  1.54$^{+0.17}_{-0.14}$   \\
GRB 050824  &  PC  &  0.90$^{+0.15}_{-0.15}$  &  1.76$^{+0.25}_{-0.41}$   \\
\hline
\end{tabular}
\end{center}
\end{table*}
\clearpage

%\addtocounter{table}{-1}
\begin{table*}
\begin{center}
\begin{tabular}{lccccccc}
\hline
GRB  & Mode & $\beta_1$ & $\beta_2$ & $\beta_3$ & $\beta_4$ & $\beta_5$ & $\beta_6$  \\
\hline
GRB 060211B  &  WT  &  1.23$^{+0.84}_{-0.38}$   \\
&  PC  &  0.70$^{+0.21}_{-0.25}$  &  0.88$^{+0.83}_{-0.61}$   \\
GRB 060218  &  WT  &  0.7474$^{+0.0102}_{-0.0099}$   \\
&  PC  &  1.91$^{+0.22}_{-0.21}$  &  4.13$^{+0.78}_{-0.17}$   \\
GRB 060219  &  PC  &  2.09$^{+1.10}_{-0.87}$  &  2.14$^{+0.26}_{-0.43}$  &  2.33$^{+0.81}_{-0.59}$   \\
GRB 060223A  &  WT    &  &  0.88$^{+0.42}_{-0.36}$   \\
&  PC  &  0.86$^{+0.32}_{-0.14}$  &  1.12$^{+0.18}_{-0.24}$   \\
\hline
\end{tabular}
\end{center}
\end{table*}
\clearpage

%\addtocounter{table}{-1}
\begin{table*}
\begin{center}
\begin{tabular}{lccccccc}
\hline
GRB  & Mode & $\beta_1$ & $\beta_2$ & $\beta_3$ & $\beta_4$ & $\beta_5$ & $\beta_6$  \\
\hline
GRB 060729  &  WT  &  2.06$^{+0.10}_{-0.10}$  &  2.96$^{+0.18}_{-0.16}$   \\
&  PC    &  &  1.96$^{+0.28}_{-0.28}$  &  1.060$^{+0.045}_{-0.044}$  &  1.088$^{+0.030}_{-0.041}$  &  0.867$^{+0.087}_{-0.120}$   \\
GRB 060801  &  PC  &  0.59$^{+0.30}_{-0.29}$  &  0.70$^{+0.26}_{-0.23}$  &  5.3$^{+2.4}_{-2.6}$   \\
GRB 060804  &  PC  &  1.09$^{+0.21}_{-0.19}$  &  1.27$^{+0.18}_{-0.34}$   \\
GRB 060805A  &  PC  &  1.28$^{+0.45}_{-0.39}$  &  1.17$^{+0.36}_{-0.59}$   \\
GRB 060807  &  PC  &  0.95$^{+0.56}_{-0.34}$  &  0.91$^{+0.21}_{-0.20}$  &  1.17$^{+0.12}_{-0.11}$  &  1.80$^{+0.27}_{-0.35}$   \\
GRB 060813  &  WT  &  0.884$^{+0.078}_{-0.078}$   \\
\hline
\end{tabular}
\end{center}
\end{table*}
\clearpage

%\addtocounter{table}{-1}
\begin{table*}
\begin{center}
\begin{tabular}{lccccccc}
\hline
GRB  & Mode & $\beta_1$ & $\beta_2$ & $\beta_3$ & $\beta_4$ & $\beta_5$ & $\beta_6$  \\
\hline
GRB 070110  &  WT  &  0.906$^{+0.090}_{-0.086}$  &  1.25$^{+0.21}_{-0.17}$   \\
&  PC  &  1.08$^{+0.23}_{-0.19}$  &  0.81$^{+0.12}_{-0.11}$  &  1.124$^{+0.076}_{-0.073}$  &  1.13$^{+0.18}_{-0.15}$  &  1.16$^{+0.20}_{-0.16}$   \\
GRB 070125  &  PC  &  1.14$^{+0.21}_{-0.23}$  &  1.17$^{+0.25}_{-0.14}$   \\
GRB 070129  &  WT  &  0.862$^{+0.027}_{-0.027}$   \\
&  PC  &  1.46$^{+0.11}_{-0.11}$  &  1.53$^{+0.28}_{-0.27}$  &  1.29$^{+0.13}_{-0.16}$  &  1.65$^{+1.26}_{-0.59}$   \\
GRB 070208  &  PC  &  1.35$^{+0.44}_{-0.26}$  &  1.26$^{+0.32}_{-0.31}$  &  1.17$^{+0.26}_{-0.37}$   \\
GRB 070219  &  WT  &  1.1$^{+4.4}_{-2.0}$   \\
\hline
\end{tabular}
\end{center}
\end{table*}
\clearpage

%\addtocounter{table}{-1}
\begin{table*}
\begin{center}
\begin{tabular}{lccccccc}
\hline
GRB  & Mode & $\beta_1$ & $\beta_2$ & $\beta_3$ & $\beta_4$ & $\beta_5$ & $\beta_6$  \\
\hline
GRB 070802  &  PC  &  0.79$^{+0.19}_{-0.17}$  &  1.22$^{+0.43}_{-0.41}$  &  1.43$^{+0.42}_{-0.43}$   \\
GRB 070808  &  WT  &  0.61$^{+0.66}_{-0.56}$   \\
&  PC  &  0.32$^{+0.44}_{-0.40}$  &  1.58$^{+0.42}_{-0.34}$   \\
GRB 070810A  &  PC  &  1.11$^{+0.36}_{-0.33}$  &  1.15$^{+0.12}_{-0.15}$   \\
GRB 071001  &  WT  &  1.95$^{+0.24}_{-0.22}$   \\
&  PC  &  1.52$^{+0.18}_{-0.24}$  &  1.35$^{+0.58}_{-0.29}$   \\
GRB 071010A  &  PC  &  1.96$^{+0.36}_{-0.38}$  &  1.40$^{+0.40}_{-0.33}$   \\
\hline
\end{tabular}
\end{center}
\end{table*}
\clearpage

\begin{table*}
\begin{center}
\begin{tabular}{lccccccc}
\hline
GRB  & Mode & $\beta_1$ & $\beta_2$ & $\beta_3$ & $\beta_4$ & $\beta_5$ & $\beta_6$  \\
\hline
GRB 080413B  &  WT  &  1.03$^{+0.10}_{-0.10}$   \\
&  PC  &  0.94$^{+0.32}_{-0.30}$  &  0.889$^{+0.051}_{-0.050}$  &  1.16$^{+0.13}_{-0.17}$   \\
GRB 080426  &  PC  &  0.63$^{+0.40}_{-0.30}$  &  1.23$^{+0.18}_{-0.27}$   \\
GRB 080430  &  WT  &  1.39$^{+0.20}_{-0.17}$   \\
&  PC  &  0.64$^{+0.23}_{-0.22}$  &  1.109$^{+0.094}_{-0.063}$  &  1.11$^{+0.11}_{-0.10}$   \\
GRB 080503  &  WT  &  0.114$^{+0.062}_{-0.062}$  &  0.724$^{+0.099}_{-0.095}$  &  1.39$^{+0.42}_{-0.37}$   \\
&  PC    &    &  &  1.21$^{+0.17}_{-0.10}$   \\
\hline
\end{tabular}
\end{center}
\end{table*}
\clearpage

%%%%%%%%%%%%%%%%%%%%%%%
%%% RES NH TABLE
%%%%%%%%%%%%%%%%%%%%%%%
\clearpage

\begin{table*}
\begin{center}
\begin{tabular}{lccccccc}
\hline
GRB  & Mode & $N_{\rm H,1}$ & $N_{\rm H,2}$ & $N_{\rm H,3}$ & $N_{\rm H,4}$ & $N_{\rm H,5}$ & $N_{\rm H,6}$  \\
\hline
GRB 050126  &  PC  &  $<10.43$  &  5.5$^{+37.1}_{-5.5}$   \\
GRB 050128  &  WT    &  &  $<123.27$   \\
&  PC  &  6.3$^{+3.0}_{-2.9}$  &  4.0$^{+1.6}_{-1.7}$   \\
GRB 050219A  &  WT  &  15.3$^{+5.2}_{-4.8}$  &  15.2$^{+8.9}_{-7.5}$   \\
&  PC    &  &  29.9$^{+28.2}_{-20.7}$   \\
GRB 050315  &  WT    &  &  112.5$^{+85.2}_{-65.7}$   \\
&  PC  &  35.8$^{+18.9}_{-17.8}$  &  79.1$^{+19.7}_{-17.1}$  &  100.0$^{+19.9}_{-14.3}$   \\
\hline
\end{tabular}
\caption{Column densities (in excess of the Galactic value) for the
time-resolved spectra, in units of $10^{20}$ \cms. $N_{\rm H,1}$ corresponds to the time during which the decay
followed $\alpha_1$ in Table~6. Spectral indices for these
spectra are given in Table~9.}
\label{tab:resnh}
\end{center}
\end{table*}

%\addtocounter{table}{-1}
\begin{table*}
\begin{center}
\begin{tabular}{lccccccc}
\hline
GRB  & Mode & $N_{\rm H,1}$ & $N_{\rm H,2}$ & $N_{\rm H,3}$ & $N_{\rm H,4}$ & $N_{\rm H,5}$ & $N_{\rm H,6}$  \\
\hline
GRB 050819  &  WT  &  $<4.12$   \\
&  PC  &  $<4.11$  &  14.0$^{+10.1}_{-6.5}$   \\
GRB 050820A  &  WT  &  2.7$^{+3.7}_{-2.7}$  &  $<1.29$   \\
&  PC    &  &  2.2$^{+1.8}_{-1.7}$  &  52.7$^{+9.3}_{-12.1}$   \\
GRB 050822  &  WT  &  13.0$^{+1.4}_{-1.3}$  &  $<18.67$  &  $<44.94$   \\
&  PC  &  3.6$^{+9.7}_{-3.6}$  &  3.5$^{+3.5}_{-1.6}$  &  18.8$^{+3.6}_{-1.7}$   \\
GRB 050824  &  PC  &  3.3$^{+6.8}_{-3.3}$  &  40.5$^{+13.6}_{-11.4}$   \\
\hline
\end{tabular}
\end{center}
\end{table*}
\clearpage

%\addtocounter{table}{-1}
\begin{table*}
\begin{center}
\begin{tabular}{lccccccc}
\hline
GRB  & Mode & $N_{\rm H,1}$ & $N_{\rm H,2}$ & $N_{\rm H,3}$ & $N_{\rm H,4}$ & $N_{\rm H,5}$ & $N_{\rm H,6}$  \\
\hline
GRB 060211B  &  WT  &  2.0$^{+17.6}_{-2.0}$   \\
&  PC  &  2.6$^{+10.8}_{-2.6}$  &  8.3$^{+29.2}_{-8.3}$   \\
GRB 060218  &  WT  &  26.34$^{+0.44}_{-0.44}$   \\
&  PC  &  22.1$^{+5.4}_{-5.0}$  &  62.0$^{+15.0}_{-7.4}$   \\
GRB 060219  &  PC  &  14.6$^{+19.3}_{-14.3}$  &  39.8$^{+12.2}_{-11.5}$  &  34.6$^{+15.7}_{-9.5}$   \\
GRB 060223A  &  WT    &  &  605.4$^{+1022.8}_{-506.2}$   \\
&  PC  &  56.1$^{+313.1}_{-56.1}$  &  319.2$^{+284.3}_{-244.8}$   \\
\hline
\end{tabular}
\end{center}
\end{table*}
\clearpage

%\addtocounter{table}{-1}
\begin{table*}
\begin{center}
\begin{tabular}{lccccccc}
\hline
GRB  & Mode & $N_{\rm H,1}$ & $N_{\rm H,2}$ & $N_{\rm H,3}$ & $N_{\rm H,4}$ & $N_{\rm H,5}$ & $N_{\rm H,6}$  \\
\hline
GRB 060729  &  WT  &  40.1$^{+4.0}_{-3.7}$  &  18.8$^{+3.7}_{-3.4}$   \\
&  PC    &  &  $<3.76$  &  12.8$^{+2.0}_{-1.9}$  &  13.0$^{+1.3}_{-1.6}$  &  16.7$^{+5.8}_{-8.2}$   \\
GRB 060801  &  PC  &  48.0$^{+42.0}_{-33.3}$  &  7.8$^{+22.0}_{-7.8}$  &  $(1.28^{+2.97}_{-0.64}) \tim{3}$   \\
GRB 060804  &  PC  &  $<7.21$  &  8.2$^{+9.5}_{-8.2}$   \\
GRB 060805A  &  PC  &  5.1$^{+7.5}_{-2.9}$  &  4.1$^{+9.8}_{-4.1}$   \\
GRB 060807  &  PC  &  1.6$^{+10.9}_{-1.6}$  &  10.9$^{+5.4}_{-4.9}$  &  13.3$^{+2.9}_{-2.6}$  &  17.0$^{+8.0}_{-11.9}$   \\
GRB 060813  &  WT  &  8.6$^{+3.4}_{-3.3}$   \\
\hline
\end{tabular}
\end{center}
\end{table*}
\clearpage

%\addtocounter{table}{-1}
\begin{table*}
\begin{center}
\begin{tabular}{lccccccc}
\hline
GRB  & Mode & $N_{\rm H,1}$ & $N_{\rm H,2}$ & $N_{\rm H,3}$ & $N_{\rm H,4}$ & $N_{\rm H,5}$ & $N_{\rm H,6}$  \\
\hline
GRB 070110  &  WT  &  10.6$^{+17.4}_{-10.6}$  &  $<9.18$   \\
&  PC  &  $<12.20$  &  $<11.15$  &  13.8$^{+14.1}_{-13.3}$  &  7.8$^{+32.2}_{-7.8}$  &  17.3$^{+24.0}_{-14.3}$   \\
GRB 070125  &  PC  &  5.8$^{+2.8}_{-4.9}$  &  3.5$^{+3.6}_{-3.5}$   \\
GRB 070129  &  WT  &  11.66$^{+0.77}_{-0.76}$   \\
&  PC  &  $<2.31$  &  16.3$^{+6.9}_{-6.4}$  &  14.2$^{+1.9}_{-4.0}$  &  11.5$^{+12.5}_{-11.5}$   \\
GRB 070208  &  PC  &  48.9$^{+40.2}_{-24.5}$  &  66.1$^{+36.3}_{-30.9}$  &  89.6$^{+38.0}_{-43.9}$   \\
GRB 070219  &  WT  &  23.6$^{+110.6}_{-23.6}$   \\
\hline
\end{tabular}
\end{center}
\end{table*}
\clearpage

%\addtocounter{table}{-1}
\begin{table*}
\begin{center}
\begin{tabular}{lccccccc}
\hline
GRB  & Mode & $N_{\rm H,1}$ & $N_{\rm H,2}$ & $N_{\rm H,3}$ & $N_{\rm H,4}$ & $N_{\rm H,5}$ & $N_{\rm H,6}$  \\
\hline
GRB 070802  &  PC  &  $<33.71$  &  161.6$^{+150.9}_{-116.6}$  &  162.5$^{+351.7}_{-81.5}$   \\
GRB 070808  &  WT  &  79.7$^{+53.2}_{-39.6}$   \\
&  PC  &  57.8$^{+38.9}_{-29.4}$  &  95.0$^{+24.5}_{-22.2}$   \\
GRB 070810A  &  PC  &  49.5$^{+72.6}_{-49.5}$  &  74.0$^{+32.9}_{-35.3}$   \\
GRB 071001  &  WT  &  $<5.06$   \\
&  PC  &  $<3.07$  &  $<30.90$   \\
GRB 071010A  &  PC  &  44.6$^{+14.0}_{-12.9}$  &  48.4$^{+13.2}_{-30.1}$   \\
\hline
\end{tabular}
\end{center}
\end{table*}
\clearpage

\begin{table*}
\begin{center}
\begin{tabular}{lccccccc}
\hline
GRB  & Mode & $N_{\rm H,1}$ & $N_{\rm H,2}$ & $N_{\rm H,3}$ & $N_{\rm H,4}$ & $N_{\rm H,5}$ & $N_{\rm H,6}$  \\
\hline
GRB 080413B  &  WT  &  27.7$^{+9.2}_{-8.6}$   \\
&  PC  &  21.4$^{+27.8}_{-21.4}$  &  23.4$^{+4.6}_{-4.5}$  &  38.2$^{+15.0}_{-19.5}$   \\
GRB 080426  &  PC  &  $<15.79$  &  21.4$^{+5.2}_{-12.5}$   \\
GRB 080430  &  WT  &  13.4$^{+6.9}_{-6.3}$   \\
&  PC  &  7.9$^{+13.6}_{-7.9}$  &  36.1$^{+6.4}_{-3.3}$  &  43.5$^{+7.8}_{-9.5}$   \\
GRB 080503  &  WT  &  2.5$^{+2.0}_{-1.9}$  &  8.6$^{+2.8}_{-2.6}$  &  11.5$^{+8.9}_{-7.5}$   \\
&  PC    &    &  &  $<1.09$   \\
\hline
\end{tabular}
\end{center}
\end{table*}

\clearpage

\begin{table*}
\begin{center}
\begin{tabular}{ll}
\hline
Type of burst  & Bursts \\
\hline
No breaks  &  050408, 050520, 051021A, 051028, 051211B, 060121, 060123, 060805B, 060901, 060928\\
  &  061025, 061122, 070309, 070724B, 070925, 071104, 080120, 080229B, 080405, 080507\\
  &  080514B, 080603A, 080613A, 080723B, 041223, 050124, 050215B, 050219B, 050223, 050326\\
  &  050412, 050509A, 050603, 050701, 050827,  {\bf 050908},  {\bf 050915A}, 051006, 051111\\
  &  051117B, 051210, 051221B, 060110,  {\bf 060116}, 060203, 060323, 060421, 060505, 060515\\
  &  060602A, 060602B, 060717, 060825, 060919, 060923B, 061006, 061110B, 061210,  {\bf 061222B}\\
  &  070227,  {\bf 070318},  {\bf 070330}, 070411, 070506, 070509, 070531, 070611, 070612B, 070714A\\
\hline
\end{tabular}
\caption{Classification of GRB light curves. Bursts in bold are those for which
a canonical light curve would be discernable. See text for details.}
\label{tab:types}
\end{center}
\end{table*}

\end{document}